\documentclass[prc,reprint, superscriptaddress,twocolumn]{revtex4-1}
\usepackage{amsmath,amssymb}
\usepackage[colorlinks=true,citecolor=blue,urlcolor=blue,linkcolor=blue]{hyperref}
\usepackage{graphicx}
\usepackage{color}

\begin{document}

\title{\boldmath  Nucleon-$\alpha$ Scattering and Resonances  in $^{5}$He and $^5$Li\\ with JISP16 
and Daejeon16 $NN$ Interactions}
\author{A. M. Shirokov}
\affiliation{Skobeltsyn Institute of Nuclear Physics, Lomonosov Moscow State University, 
 Moscow 119991, Russia}
\affiliation{Department of Physics and Astronomy, Iowa State University, Ames, Iowa 50011, USA}
\affiliation{Department of Physics, Pacific National University, Khabarovsk 680035, Russia}
\author{A. I. Mazur}
\affiliation{Department of Physics, Pacific National University, Khabarovsk 680035, Russia}
\author{I. A. Mazur}
\affiliation{Department of Physics, Pacific National University, Khabarovsk 680035, Russia}
\author{E. A. Mazur}
\affiliation{Department of Physics, Pacific National University, Khabarovsk 680035, Russia}
\author{I. J. Shin}
\affiliation{Rare Isotope Science Project, Institute for Basic Science, Daejeon 305-811, Korea}
\author{Y. Kim}
\affiliation{Rare Isotope Science Project, Institute for Basic Science, Daejeon 305-811, Korea}
\author{L. D. Blokhintsev}
\affiliation{Skobeltsyn Institute of Nuclear Physics, Lomonosov Moscow State University, 
 Moscow 119991, Russia}
\affiliation{Department of Physics, Pacific National University, Khabarovsk 680035, Russia}
\author{J. P. Vary}
\affiliation{Department of Physics and Astronomy, Iowa State University, Ames, Iowa 50011, USA}

й
\begin{abstract}
The SS-HORSE approach to analysis of resonant states is generalized to the case of charged
particle scattering utilizing analytical properties of partial scattering amplitudes and applied to the study
of resonant states in the $^{5}$Li nucleus and non-resonant $s$-wave proton-$\alpha$ scattering within the
no-core shell model using the JISP16 and Daejeon16 $NN$ interactions. We present also the results of
calculations of neutron-$\alpha$ scattering and resonances in the $^{5}$He nucleus with Daejeon16 
and compare with results published previously using JISP16.
\end{abstract}


\maketitle

\section{Introduction}

There is considerable  progress in developing {\em ab initio} methods for studying nuclear structure~\cite{Leidemann} 
based on a rapid development of supercomputer facilities and recent advances in the utilization of 
high-performance computing systems. In particular, modern {\em ab initio} approaches, such as
the Green's Function Monte Carlo (GFMC)~\cite{GFMC-3}, the Hyperspherical 
expansion~\cite{Leidemann}, the No-Core Shell Model 
(NCSM)~\cite{NCSM-2}, the Coupled-Cluster Theory~\cite{CCM-1,CCM-2}, and the
Nuclear Lattice Effective Field Theory~\cite{EFT-Lee,EFT-Epel}
are able to reproduce properties of  atomic nuclei with mass up to $A = 16$ 
and selected heavier nuclear systems around closed shells.

Within the NCSM as well as within other 
variational approaches utilizing the harmonic oscillator basis,
the calculation of nuclear ground states and other bound states
starts conventionally from estimating the 
dependence of the 
energy~$E_{\nu}(\hbar\Omega)$ of the bound state~$\nu$ in some model space on the
harmonic oscillator frequency~$\hbar\Omega$. 
The minimum of~$E_{\nu}(\hbar\Omega)$ is correlated with the energy of the state~$\nu$. The 
convergence of calculations and accuracy of the energy prediction is estimated by comparing with the 
results obtained in neighboring model spaces. To improve the accuracy of theoretical predictions, various 
extrapolation techniques have been suggested 
recently~\cite{NCSM-Extra, Extra-Coon,CoonNTSE12,Extra-Furn, Extra-More, Extra-Kruse, Extra-Saaf, Extra-Furn2, Extra-Konig, Extra-Furn3, Extra-Wendt, Extra-Coon2,6Li-2017,neural} 
which make it possible to estimate the binding energies in the complete infinite 
basis space. 
The studies of extrapolations to the infinite model spaces reveal general trends of convergence patterns
of variational calculations with the harmonic oscillator basis, in the
shell model calculations in particular.

An extension of the {\em ab initio} methods to the studies of the continuum spectrum and nuclear
reactions is one of the mainstreams of modern nuclear theory. A remarkable success in developing 
 the
{\em ab initio} reaction theory was achieved in  few-body physics where exact Faddeev and 
Faddeev--Yakubovsky equations~\cite{Merk-Fad} or the AGS method~\cite{Alt} are nowadays
routinely used for calculating various few-body reactions.

The most important breakthrough in developing {\em ab initio} theory of nuclear reactions in systems
with total number of nucleons~$A>4$ was achieved by combining NCSM and Resonating Group Method
(RGM); the resulting approaches are conventionally referred to as NCSM/RGM and the
No-Core Shell Model with Continuum
\mbox{(NCSMC)}~\cite{NCSM_RGM_PRC82,NCSM_RGM_JPG36,NCSM_RGM_PhScr91,NCSM-2}.
It is also worth noting the Lorentz integral transform approach to nuclear reactions with
electromagnetic probes~\cite{Efros,Leidemann} and the GFMC calculations of elastic~$n\alpha$
scattering~\cite{Nollett}. Nuclear resonances can be also studied within the No-core Gamow Shell Model 
(NCGSM)~\cite{NCGSM}.

Both NCGSM and NCSM/RGM complicate essentially the shell model calculations.
A conventional belief is that the energies of shell model states in the continuum should be associated 
with the resonance energies. It was shown however in Refs.~\cite{naPRC, naAMIS} that the energies 
of shell model states may appear well above the energies of resonant states, especially for broad 
resonances. Moreover, the analysis of Refs.~\cite{naPRC, naAMIS} clearly demonstrated that the 
shell model should also generate some states in a non-resonant nuclear continuum. 
In Refs.~\cite{SSHORSEPRC,SSHORSE,PEPAamN,BMMS,BMMS_Z} we suggested 
an SS-HORSE approach which provides an interpretation of the shell model states in the continuum
and makes it possible to deduce resonance energies and widths or low-energy non-resonant phase shifts directly
from shell-model results  without introducing additional Berggren basis states as in NCGSM or 
additional 
calculations as in the NCSM/RGM and NCSMC approaches.

The SS-HORSE approach is based on a simple analysis of the~$\hbar\Omega$ and basis size 
dependencies of the results of standard variational
shell-model calculations. We have successfully applied it to extracting  resonance
energies and widths in~$n\alpha$
scattering as well as non-resonant~$n\alpha$ elastic scattering phase 
shifts~\cite{SSHORSEPRC,SSHORSE} from the NCSM
calculations of $^{5}$He and $^{4}$He nuclei with the JISP16 $NN$ interaction~\cite{PLB644}.
To describe democratic decays~\cite{Jib-Krupennikova,JibutiEChAYa} of few-nucleon systems,
we developed a hyperspherical extension of the SS-HORSE method~\cite{tetra-ntse2014,PRL4n}.
An application of this extended SS-HORSE approach 
to the study of the four-neutron system (tetraneutron)~\cite{tetra-ntse2014,PRL4n,Igor-here} predicted
for the first time a low-energy tetraneutron resonance consistent
with a recent experiment~\cite{4nExp} with soft realistic $NN$ interactions like
JISP16~\cite{PLB644}, 
Daejeon16~\cite{Daejeon16} 
and SRG-softened 
chiral effective
field theory ($\chi$EFT) 
$NN$ interaction of Ref.~\cite{IdN3LO}. 

In this contribution, we discuss an extension of the SS-HORSE method to the case of 
charged particle scattering. The SS-HORSE technique provides the $S$-matrix or scattering
phase shifts in some energy interval  above the threshold where the shell model calculations
generate eigenstates with various~$\hbar\Omega$ values and various basis truncations. Next we
parametrize the $S$-matrix to obtain it in a wider energy interval and to locate its poles associated
with resonances. We have shown~\cite{SSHORSEPRC,SSHORSE} that this
parametrization should provide a correct description of low-energy phase shifts.  The phase shift
parametrization utilized in Refs.~\cite{SSHORSEPRC,SSHORSE,PEPAamN} was derived from
the symmetry properties of the $S$-matrix. However, due to the long-range Coulomb interaction in the case
of charged particle scattering, the analytical properties of the $S$-matrix become much more
complicated and cannot be used for its low-energy parametrization. In Ref.~\cite{BMMS} we suggested
a version of the SS-HORSE approach which utilizes the phase shift parametrization based on
analytical properties of the partial-wave scattering amplitude. In the case of  charged particle scattering,
instead of the partial-wave scattering amplitude, one can use the so-called
renormalized Coulomb-nuclear amplitude~\cite{HOT,Haeringen} which has similar analytical properties.
This opens a route to the generalization of the SS-HORSE method to the case of the
charged particle scattering proposed in Ref.~\cite{BMMS_Z} where we have verified this approach
using a model problem of scattering of particles interacting by the Coulomb and a short-range
potential. To calculate the Coulomb-nuclear phase shifts, we make use of the version of the
HORSE formalism suggested in Ref.~\cite{Bang-m} and utilized later in our studies of
Refs.~\cite{naPRC, naAMIS}.

In this contribution we present the results of SS-HORSE calculations of proton-$\alpha$ resonant and
non-resonant scattering phase shifts based on the {\em ab initio} NCSM results 
for $^{5}$Li and $^{4}$He nuclei
obtained with the JISP16~\cite{PLB644} and a
newer Daejeon16~\cite{Daejeon16} $NN$ interaction derived from a $\chi$EFT 
inter-nucleon potential and
better fitted to the description of light nuclei than JISP16. We search for the $S$-matrix poles to
evaluate the  energies and widths of resonant states in the $^{5}$Li nucleus. The NCSM-SS-HORSE
calculations of the $^{5}$He resonant states have been performed with the JISP16 interaction in
Refs.~\cite{SSHORSEPRC,SSHORSE}. We present here also the results of the NCSM-SS-HORSE
$^{5}$He resonant state calculations with the Daejeon16 to complete the studies of the
nucleon-$\alpha$ resonances with the realistic JISP16 and Daejeon16 $NN$ potentials.
The previous  {\em ab initio} analyses of nucleon-$\alpha$ resonances with various modern
realistic inter-nucleon interactions were performed in Ref.~\cite{Nollett} within the GFMC,
within the NCGSM in Ref.~\cite{NCGSM}, within the Coupled-Cluster Theory with
Berggren basis in Ref.~\cite{CCn-alpha},
and in
\mbox{Refs.~\cite{NCSM_RGM_PRC79,NCSM_RGM_PRC88,NCSM_RGM_PRC90,NCSM_RGM_PRC82}} 
within the NCSM/RGM. {We note also a recent paper of R.~Lazauskas~\cite{Lazauskas} where
the $n\alpha$ scattering was studied within a five-body Faddeev--Yakubovsky approach.}

\section{SS-HORSE method for channels with neutral and charged particles}

\subsection{General formulae}
The SS-HORSE approach relies on the $J$-matrix formalism in quantum scattering theory.

Originally, the $J$-matrix formalism was developed
in atomic physics~\cite{HeYa74}; therefore, the so-called
Laguerre basis was naturally used within this approach. A generalization of this formalism
utilizing either the Laguerre or the harmonic oscillator bases was suggested in Ref.~\cite{YaFi-m}.
Later the harmonic-oscillator version of the $J$-matrix method was independently rediscovered by
Kiev (G.~F.~Filippov and collaborators)~\cite{Fillip}  and Moscow 
(Yu.~F.~Smirnov and collaborators)~\cite{NeSm} 
groups. The $J$-matrix with oscillator basis is sometimes also referred to as an {\em Algebraic
Version of RGM}~\cite{Fillip} or as  a {\em Harmonic Oscillator Representation of Scattering Equations}
(HORSE)~\cite{Bang-m}. We use here a generalization of the HORSE formalism to the case of
charged particle scattering proposed in Ref.~\cite{Bang-m}.

Within the HORSE approach, the basis function space is split into internal and external regions. 
In the internal region which includes the basis states with oscillator  quanta $N\leq \mathbb N$,
the Hamiltonian completely accounts for the kinetic and potential energies. The internal region
can be naturally associated with the shell model basis space. In the external
region, the Hamiltonian accounts for the relative kinetic energy of the colliding particles (and for 
their internal Hamiltonians if needed) only and its matrix takes a form of an infinite tridiagonal
matrix of the kinetic-energy operator (plus the sum of eigenenergies of the  colliding particles at
the diagonal if 
they have an internal structure). The external region clearly 
represents the scattering channel under consideration. If the eigenenergies~$E_{\nu}$,
$\nu=0$, 1,~...\,, and the
respective eigenvectors of the Hamiltonian matrix in the internal region are known, one can
easily calculate the $S$-matrix,  phase shifts and other parameters characterizing the
scattering process (see, e.\:g., Refs.~\cite{YaFi-m,Bang-m,Svesh-mem,TMF-m}).

An interesting
feature peculiar to the $J$-matrix method was highlighted
as far back as 1974~\cite{HeYa74}. The point is that, at the
energies coinciding with the eigenvalues~$E_{\nu}$ of the Hamiltonian
matrix in the internal region, the matching condition of
the $J$-matrix method becomes substantially simpler while the accuracy of the $S$-matrix
and phase shift description  at these energies is much better than at the energies away 
from the eigenvalues~$E_{\nu}$~\cite{YaAb93,Ya2013,BMMS_Z}.  Taking an advantage
of this feature, H.~Yamani~\cite{Ya2013} was able to construct an
analytic continuation to the complex energy plane within
the $R$-matrix method and to obtain
accurate estimates for the energies and widths of
resonant states.


The Single-State HORSE (SS-HORSE) method suggested in 
Refs.~\cite{SSHORSEPRC,SSHORSE,PEPAamN} also benefits from the improved accuracy
of the HORSE approach at the eigenstates of the Hamiltonian matrix truncated to the
internal region of the whole basis space. In the case of scattering of uncharged particles interacting
by a short-range potential, the phase shifts~$\delta_{l} (E_{\nu})$ 
in the partial wave with the orbital momentum~$l$
at the eigenenergies~$E_{\nu}$ of the internal Hamiltonian
matrix are given by~\cite{SSHORSEPRC,SSHORSE,PEPAamN} 
\begin{equation}
 \label{SSJM_phase}
\tan{\delta_{l} (E_{\nu})}  = -\frac{S_{{\mathbb N}+2,l}(E_{\nu})}{C_{{\mathbb N}+2,l}(E_{\nu})} .
\end{equation}
Here $S_{N,l}(E)$ and $C_{N,l}(E)$ are respectively regular and irregular solutions of the free
Hamiltonian at energy~$E$ in the oscillator representation for which analytical expressions
can be found in Refs.~\cite{YaFi-m,Bang-m,Svesh-mem,TMF-m}. Varying the oscillator 
spacing~$\hbar\Omega$ and the truncation boundary~$\mathbb N$ of the internal oscillator
basis subspace, we obtain a variation of 
the eigenenergy~$E_{\nu}$ of the truncated
Hamiltonian matrix over an 
energy interval and obtain the phase shifts~$\delta_{l} (E)$
in that energy interval by means of Eq.~\eqref{SSJM_phase}. Next, we parametrize the
phase shifts~$\delta_{l} (E)$ as discussed in the next subsection to have the phase shifts
and the $S$-matrix in a wider energy interval which makes it possible to locate the $S$-matrix poles.

In the case of scattering in the channels with two charged particles, 
 following the ideas of
Ref.~\cite{Bang-m}, we formally cut the Coulomb interaction at the distance $r=b$. As shown in 
Ref.~\cite{BMMS_Z}, 
an optimal value of the Coulomb cutoff distance~$b$ is the so-called 
{\em natural channel radius}~$b_0$~\cite{Bang-m},
\begin{equation} 
\label{brcl}
b= b_0\equiv r_{\mathbb N+2,l}^{cl} = 2r_0\sqrt{\mathbb N/2+7/4}\, ,
\end{equation}
i.\:e., $b$ is equivalent to the classical turning point~$r_{\mathbb N+2,l}^{cl}$ of the first oscillator 
function~$R_{\mathbb N+2,l}(r)$ in the external region of the basis space. The 
parameter~$r_0=\sqrt{\hbar/(\mu\Omega)}$ entering Eq.~\eqref{brcl} is the oscillator radius and~$\mu$
is the reduced mass in the channel under consideration. With this choice of the 
Coulomb cutoff distance~$b$, the  elements of the Hamiltonian matrix in the internal region are
insensitive to the cut of the Coulomb interaction. Therefore the 
shell model Hamiltonian matrix elements 
 in the internal region can be calculated without any modification of the Coulomb interaction
between the nucleons. The scattering phase shifts~$\delta_{l}^{aux}$ of the auxiliary Hamiltonian with the cutoff
Coulomb interaction can be calculated using the standard  HORSE or SS-HORSE technique, e.\:g.,
with the help of Eq.~\eqref{SSJM_phase}. To deduce an expression for the Coulomb-nuclear phase shifts~$\delta_{l}$,
one should match at the distance~$b$ the plane-wave asymptotics of the auxiliary  Hamiltonian
wave functions with Coulomb-distorted wave function asymptotics. As a result, we get the following
SS-HORSE expression for the 
Coulomb-nuclear  phase shifts~$\delta_{l} (E_{\nu})$ 
at the eigenenergies~$E_{\nu}$ of the internal Hamiltonian
matrix~\cite{BMMS_Z}:
\begin{multline}
 \label{SSJM_phZ}
\tan{\delta_{l} \big(E_{\nu}\big)}  \\
=-\frac{S_{\mathbb N+2,l}\big(E_{\nu}\big)\,W_b({n}_{l},F_{l})+
C_{\mathbb N+2,l}(E_{\nu})\,W_b({j}_{l},F_{l})}{S_{\mathbb N+2,l}\big(E_{\nu}\big)\,W_b({n}_{l},G_{l})+
C_{\mathbb N+2,l}\big(E_{\nu}\big)\,W_b({j}_{l},G_{l})} .
\end{multline}
Here $j_{l}\equiv j_{l}(kr)$ and~$n_{l}\equiv n_{l}(kr)$ are respectively the spherical Bessel and Neumann functions~\cite{DLMF}
while $F_{l}\equiv F_{l}(\eta,kr)$ and $G_{l}\equiv G_{l}(\eta,kr)$ are respectively 
the regular and irregular
Coulomb functions~\cite{DLMF}; $k$ is the relative motion momentum; 
$\eta=Z_1Z_2e^2\mu /(\hbar^{2}k)$ is the
Sommerfeld parameter; the quasi-Wronskian
  \begin{equation}
  W_b(\phi,\chi)= \left. \left(\frac{d\phi}{dr}\,\chi\,-\,\phi\,\frac{d\chi}{dr} \right)\!
 \right|_{r=b}\! .
  \label{qwron}
  \end{equation}

As in the case of neutral particle scattering, we obtain the Coulomb-nuclear phase shifts~$\delta_{l}(E)$
in some energy interval by varying the internal region boundary~$\mathbb N$ and the
oscillator 
spacing~$\hbar\Omega$, then 
parametrize the phase shifts to have them
in a wider energy interval. However the phase shift parametrization is more complicated for
channels with charged colliding particles as discussed below.

An important scaling property of variational calculations with the harmonic oscillator basis
was revealed in Refs.~\cite{Extra-Coon,CoonNTSE12}: 
the converging variational
eigenenergies~$E_{\nu}$ do not depend on~$\hbar\Omega$ and~$\mathbb N$ independently
but only through a scaling variable
\begin{gather}
s=\frac{\hbar\Omega}{{\mathbb N}+7/2}.
\label{scales}
\end{gather}
This scaling property was initially proposed in Refs.~\cite{Extra-Coon,CoonNTSE12} for the
bound states. We have extended the scaling to the case of variational calculations 
with the harmonic oscillator basis of the unbound states~\cite{SSHORSEPRC,SSHORSE}
within the SS-HORSE approach. The SS-HORSE extension to the case of charged particle
scattering discussed here can be used to demonstrate that the
long-range Coulomb interaction does not destroy the
scaling property of the unbound states (see Ref.~\cite{BMMS_Z} for details).

\subsection{Phase shift parametrization}
The total partial-wave scattering amplitude 
in the case of Coulomb and short-range
interactions has the form of the sum of the purely
Coulomb,~$ f_l^{C}(k)$, and Coulomb-nuclear,~$f_l^{NC}(k)$, amplitudes~\cite{Newton},
\begin{equation}
  \label{fl}
  f_l(k) = f_l^{C}(k)+f_l^{NC}(k),
\end{equation}
which, in turn, are related to the purely Coulomb, $\sigma_l=\arg \Gamma (1+l+i \eta)$, 
and Coulomb-nuclear phase shifts,
$\delta_l$, as
\begin{equation}
  \label{flC}
   f_l^{C}(k)=\frac{\exp{(2i\sigma_l)}-1}{2ik},
\end{equation}
\begin{equation}
  \label{flNC}
   f_l^{NC}(k)=\exp{(2i\sigma_l)}\,\frac{\exp{(2i\delta_l)}-1}{2ik}.
\end{equation}

Analytic properties of the Coulomb-nuclear amplitude~$f_l^{NC}(k)$
in the complex momentum plane  differ from the analytic properties of
the scattering amplitude for neutral particles. However,
the {\em renormalized Coulomb-nuclear amplitude}~\cite{HOT, Haeringen},
\begin{equation}
	{\widetilde f}_l(E)=\frac{\exp{(2i\delta_l)}-1}{2ik}\cdot \frac{\exp{(2\pi\eta)}-1}{2\pi\eta}
	\, c_{l\eta},
      \label{B-5}
\end{equation}\vspace{-.9ex}
where\vspace{-1.3ex}
\begin{gather}
\label{cleta}
c_{l\eta} = \prod _{n=1}^l(1+\eta^2/n^2)^{-1} \ \  (l>0),\ \ \ \ \  c_{0\eta}=1,
\end{gather}
is  identical in analytic properties on the real momentum
axis with the scattering amplitude for neutral particles.
In particular, the renormalized amplitude can
be expressed~\cite {HOT,Haeringen}  
\begin{gather}
\widetilde{f}_{l}(E)=\frac{k^{2l}}{\widetilde{K}_{l}(E)-2\eta k^{2l+1}H(\eta)(c_{l\eta})^{-1}} 
\label{Bl6}
\end{gather}
in terms of the {\em Coulomb-modified
effective-range function}~\cite {HOT,Haeringen}  
\begin{multline}
\widetilde{K}_{l}(E) =  k^{2l+1}(c_{l\eta})^{-1}  \\[.8ex] 
\times
\!\left\{  \frac{2\pi\eta}{\exp{(2\pi\eta)}-1}
 \left[\cot\delta_l(E)-i\right]+2\eta H(\eta)
\right\}\! ,
\label{Bl7}
\end{multline}
where \vspace{-1.2ex}
\begin{gather}
H(\eta) = \Psi(i\eta)+(2i\eta)^{-1} -\ln{(i\eta)},
\end{gather}
$\Psi(z)$ is the logarithmic derivative of the~$\Gamma$ function
(digamma or~$\Psi$ function)~\cite{DLMF}, the relative motion 
energy~$E=\hbar^2k^2/(2\mu)$. 
In the absence
of Coulomb interaction ($\eta = 0$), the  Coulomb-modified effective-range function transforms into
the standard effective-range function for neutral particle scattering,
\begin{equation}
\label{Kldef}
  \widetilde{K}_l(E)=K_l(E) = k^{2l+1}\cot\delta_l,
  \end{equation}
while the renormalized amplitude becomes the conventional neutral particle scattering amplitude,\begin{equation}
	f_l(E)
	= \frac{k^{2l}}{K_l(E)-ik^{2l+1}}  .
\label{fl_def}
\end{equation}

Due to their nice analytic properties, the renormalized Coulomb-nuclear amplitude,~${\widetilde f}_l(E)$,
and the neutral particle scattering amplitude, $f_l(E)$, can be used to parametrize respectively
the Coulomb-nuclear and neutral particle scattering phase shifts ensuring their correct 
low-energy behavior. In Refs.~\cite{BMMS,BMMS_Z}, we introduced an auxiliary
 complex-valued function  embedding  resonant pole parameters in the amplitude parametrization.
 These resonant pole parameters play the role of additional fitting parameters in the phase-shift
parametrization. Here we prefer to parametrize the Coulomb-modified effective-range 
function~\eqref{Bl7} or the standard effective-range function for neutral particle scattering~\eqref{Kldef}
thus reducing the number of fit parameters. The resonant  parameters are obtained
by a numerical location of the amplitude pole as discussed below.

The Coulomb-modified effective-range function~$\widetilde{K}_{l}(E)$ 
as well as the effective-range function for neutral particle scattering~$K_l(E)$
 is real on the real axis of momentum~$k$, is regular in
the vicinity of zero, and admits an expansion in
even powers of $k$, or, equivalently, in power series of the relative motion 
energy~$E=\hbar^2k^2/(2\mu)$~\cite{HOT,Haeringen},\vspace{-1.2ex}
\begin{equation}
       \label{K_l}
	\widetilde{K}_l(E) =
	w_0+w_1 E+w_2 \hspace{1pt}E^2+ ... 
\end{equation}
The  expansion coefficients~$w_0$ and~$w_1$ are related
to the so-called scattering length~$a_l$ and effective range~$r_l$~\cite{Newton}: 
\begin{equation}
\label{w1w2alrl}
	w_0 = -\frac{1}{a_l}, \quad  w_1= \frac{r_l \mu}{\hbar^2} .
\end{equation}
We use the expansion coefficients~$w_0$, $w_1$ and $w_2$ as fit parameters for the
phase shift parametrization. Such a parametrization works well in the case of nucleon-$\alpha$
scattering but may fail in other problems. Note, as seen from Eq.~\eqref{Bl7} or Eq.~\eqref{Kldef},
the positive energies at which the phase shift takes the values of~0, $\pm \pi$, $\pm2\pi$,\,..., are the
singular points of the effective-range function. In the case of possible presence of such singular
points in the range of energies of interest for a particular problem, one should use a more elaborate
parametrization of the effective-range function, e.\:g., in the form of the Pad\'e approximant. 

\subsection{Fitting process}

In the case of neutral particle scattering, we combine Eqs.~\eqref{SSJM_phase}, \eqref{Kldef}
and~\eqref{K_l} to obtain
 \begin{equation}
	w_0+w_1E+w_2 \hspace{.8pt}E^2= -k^{2l+1}\,\frac{\displaystyle C_{{\mathbb N}+2,l}(E)}{\displaystyle S_{{\mathbb N}+2,l}(E)}.
	     \label{res_K_N_fit}
\end{equation}
In the case of charged particle scattering, we derive a more complicated equation with the
help of Eq.~\eqref{SSJM_phZ}  and~\eqref{Bl7}:
\begin{multline}
	w_0+w_1E+w_2\hspace{.8pt} E^2= 
- k^{2l+1}(c_{l\eta})^{-1} 
\left\{  \frac{2\pi\eta}{\exp{(2\pi\eta)}-1} \right. \\
\times
\left[\frac{\displaystyle S_{\mathbb N+2,l}(E)\,W_b({n}_{l},G_{l})+C_{\mathbb N+2,l}(E)\,W_b({j}_{l},G_{l})}
{\displaystyle S_{\mathbb N+2,l}(E)\,W_b({n}_{l},F_{l}) +
C_{\mathbb N+2,l}(E)\,W_b({j}_{l},F_{l})}+i\right]\\
\left.  -2\eta H(\eta) \vphantom{\frac12}\right\}\!.
	     \label{res_K_Z_fit}
\end{multline}

Let $E^{(i)}_{\nu}$, $i=1$, 2,\,...\,, $D$, be a set of the lowest ($\nu=0$) or some other 
particular eigenvalues ($\nu>0$)
of the Hamiltonian matrix truncated to the internal region of the basis space obtained with  a set 
of parameters~$({\mathbb N}^{(i)\!}, \hbar\Omega^{(i)})$, $i=1$, 2,\,...\,, $D$. 
We find energies~${\cal E}^{(i)}$ as
solutions of Eq.~\eqref{res_K_N_fit} or  Eq.~\eqref{res_K_Z_fit}
with some trial set
of the effective-range function expansion coefficients~$w_0$, $w_1$, $w_2$ for each
combination of parameters~$({\mathbb N}^{(i)\!}, \hbar\Omega^{(i)})$ [note, the oscillator
basis parameter~$\hbar\Omega$ enters definitions of functions~$S_{N,l}(E)$ and~$C_{N,l}(E)$].
The optimal set of the fit parameters~$w_0$, $w_1$, $w_2$ parametrizing the phase shifts is obtained by
minimizing the functional
\begin{equation}
 \label{ksirdyc}
         \Xi = \sqrt{\frac{1}{D} \sum_{i=1}^{D}\bigl(  E^{(i)}_{\nu}-{\cal E}^{(i)}   \bigr)^2}.
\end{equation}

With the optimal  set of the fit parameters~$w_0$, $w_1$, $w_2$ we can use 
Eq.~\eqref{res_K_N_fit} or  Eq.~\eqref{res_K_Z_fit} to obtain the~$\hbar\Omega$
dependences of the eigenenergies $E_{\nu}(\hbar\Omega)$  in any basis space~${\mathbb N}$.
Therefore Eqs.~\eqref{res_K_N_fit} and~\eqref{res_K_Z_fit} provide the extrapolation of the
variational results for unbound states to larger basis spaces.

\subsection{Resonance energy \boldmath$E_{r}$ and width $\Gamma$\label{secrespole}}
We obtain resonance energies~$E_{r}$ and widths~$\Gamma$ by a numerical location of the
$S$-matrix poles which coincide with the poles of the scattering amplitude. If the amplitude
has a resonant pole at a complex energy~$E=E_{p}$, the resonance energy~$E_{r}$ and its 
width~$\Gamma$ are related to the real and imaginary part of~$E_{p}$~\cite{Newton}:
 \begin{equation}
 \label{Epol}
 E_p=E_r-i\frac{\Gamma}{2}.
          \end{equation}
          
It follows from Eqs.~\eqref{Bl6} and~\eqref{fl_def} that locating the pole of the scattering amplitude
is equivalent to solving in the complex energy plane the equation
\begin{gather}
{\cal F}(E)\equiv\widetilde{K}_{l}(E)-2\eta k^{2l+1}H(\eta)(c_{l\eta})^{-1} =0
\label{Ampl_pol}
\end{gather}
in the case of charged particle scattering or the equation
\begin{gather}
{\cal F}(E)\equiv K_l(E)-ik^{2l+1} =0
\label{Ampl_pol_Z0}
\end{gather}
in the case of neutral particles. We can use the parametrization of functions~$\widetilde{K}_{l}(E)$
or~$K_l(E)$ in Eqs.~\eqref {Ampl_pol} and~\eqref{Ampl_pol_Z0}.
To solve these equations, we calculate the integral 
\begin{gather}
 \Upsilon  = \frac{1}{2\pi i}\oint\limits_{C} \frac{{\cal F}^{\prime}(E)}{{\cal F}(E)}\,dE
 \label{numbZero}
\end{gather}
along some closed
contour~$C$ in the complex energy plane,
where ${\cal F}^{\prime}(E) = 
\frac{d{\cal F}}{dE}$. The contour~$C$ should surround the area where we expect to have the
pole of the amplitude.
According to the 
theory of functions of a complex variable~\cite{TFKP}, the value of~$ \Upsilon$ is equal to the 
number of zeroes of the function~${\cal F}(E)$ in the area surrounded by the contour~$C$. 
 If needed, we modify
the contour~$C$ to obtain
\begin{gather}
\Upsilon  = 1. 
\label{Upsilon1}
\end{gather}
The position of the pole in the energy plane is calculated as
\begin{gather}
 E_p  = \frac{1}{2\pi i}\oint\limits_{C} E\, \frac{{\cal F}^{\prime}(E)}{{\cal F}(E)}\,dE .
 \label{valEpol}
\end{gather}

A numerical realization of the algorithm based on Eqs.~\eqref{numbZero}--\eqref{valEpol} provides
means for a fast and stable determination
of the poles of scattering amplitude.

\section{Elastic scattering of nucleons by \boldmath$\alpha$ particle\\ in the NCSM-SS-HORSE
approach}
\label{ch:Nalpha} 

We present here an application of our SS-HORSE technique to nucleon-$\alpha$ scattering phase shifts 
and resonance parameters based on {\em ab initio} many-body calculations of $^{5}$He and
$^{5}$Li nuclei within the NCSM with the realistic JISP16 and Daejeon16 $NN$ interactions. 
The NCSM calculations are performed using the code MFDn~\cite{mMFDnPCS,mMFDnCCPE} with 
basis spaces including all many-body oscillator states with excitation quanta~$N_{\max}$ ranging
from~2  up to~18 for both parities and with~$\hbar\Omega$ values ranging from~10 to~40~MeV 
in steps of~2.5~MeV.

Note,  for the NCSM-SS-HORSE analysis we need the $^5$He and $^5$Li  energies relative 
respectively to the~$n+\alpha$ and
$p + \alpha$ thresholds. Therefore from each of the $^5$He or $^5$Li NCSM odd (even) 
parity eigenenergies we 
subtract the $^4$He ground state energy obtained by the NCSM with the same $\hbar\Omega$ and the same $N_{\max}$ (with $N_{\max} -1$) excitation quanta, and in what follows these subtracted energies are referred to as the NCSM eigenenergies $E_{\nu}$. Only these $^5$He and 
$^5$Li NCSM eigenenergies relative to the respective threshold are discussed below. 

We note here that the NCSM utilizes the truncation based on
the many-body oscillator quanta $N_{\max}$ while the SS-HORSE
requires the oscillator quanta truncation~${\mathbb N}$ of the interaction
describing the relative motion of nucleon and~$\alpha$ particle. 
We relate~${\mathbb N}$ to~$N_{\max}$ as
\begin{gather}
{\mathbb N}=N_{\max}+N_{0},
\label{NbbNmaxNo}
\end{gather}
where $N_{0}=1$  is the minimal oscillator quanta in our five-body~$N\alpha$ systems. A
justification of using 
this relation for the SS-HORSE analysis is
obvious if the~$\alpha$ particle is described by the simplest four-nucleon oscillator function with excitation quanta~$N_{\max}^{\alpha}=0$.
Physically it is clear that the use of~Eq.~\eqref{NbbNmaxNo} for 
the SS-HORSE should work well also in a more general case when the~$\alpha$ particle is presented by the wave function with~$N_{\max}^{\alpha}>0$ 
due to the dominant role of the zero-quanta component in the~$\alpha$ particle wave function. 
Instead of attempting to justify algebraically the use of $N_{\max}$ within the SS-HORSE, 
we suggested in Ref.~\cite{SSHORSEPRC,SSHORSE} an {\em a posteriori} 
justification: we demonstrated  in Ref.~\cite{SSHORSEPRC,SSHORSE}  
that we obtained~$n\alpha$ phase shift parametrizations consistent with the NCSM results obtained with very different $N_{\max}$ and~$\hbar\Omega$ values; more, we were able to
predict the NCSM results with large $N_{\max}$ using the phase shift parametrizations based on the NCSM calculations with much smaller model spaces. It 
would clearly be
impossible if the use of $N_{\max}$ truncation for the SS-HORSE analysis did not work properly. We performed the same {\em a posteriori} 
analysis of our results in the present study of~nucleon-$\alpha$ scattering to ensure the justification of our approach 
though we do not present and discuss it below. Generally, the fact that the phase shifts calculated using
Eq.~\eqref{SSJM_phase} or~\eqref{SSJM_phZ} at the NCSM eigenenergies obtained with
different~$N_{\max}$ truncations form a single curve as a function of energy serves as
a confirmation of the consistency of the whole
NCSM-SS-HORSE approach and of the use  of the NCSM~$N_{\max}$ for the SS-HORSE phase
shift calculation in particular. The ranges of~$N_{\max}$ and~$\hbar\Omega$ values where this
consistency is achieved differ for different $NN$ interactions and different angular momenta
and parities. Such a consistency, which can be also interpreted as a convergence of the phase shift
calculations, is seen in the figures below to be achieved in all calculations at least at largest basis
spaces in some range of~$\hbar\Omega$ values.


\subsection{Phase shifts of resonant \boldmath$p\alpha$ scattering}

\begin{figure}[t!]
\centerline{\includegraphics[width=\columnwidth]{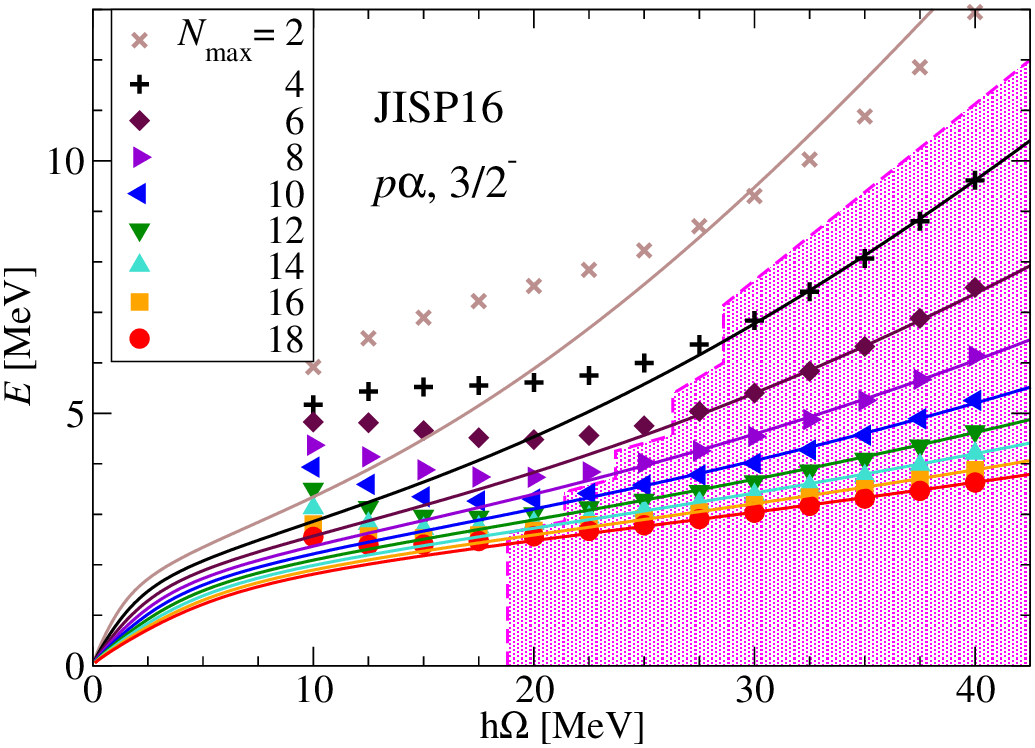}}
\caption{
The lowest $^{5}$Li $\frac32^{-}$ eigenenergies~$E^{(i)}_{0}$ relative to the~$p+\alpha$ threshold
obtained by the 
NCSM with the JISP16 $NN$ interaction 
with various~$N_{\max}$ (symbols) as functions of~$\hbar\Omega$. The shaded area shows 
the energy values selected for the SS-HORSE analysis. Solid curves are solutions 
of Eq.~\eqref{res_K_Z_fit}
for energies~$E$ with  parameters~$w_0$, $w_1$ and~$w_2$ obtained by the fit.}
\label{pa_p3_J_Ehw}      
\end{figure}

\begin{figure}[t!]
\centerline{\includegraphics[width=\columnwidth]{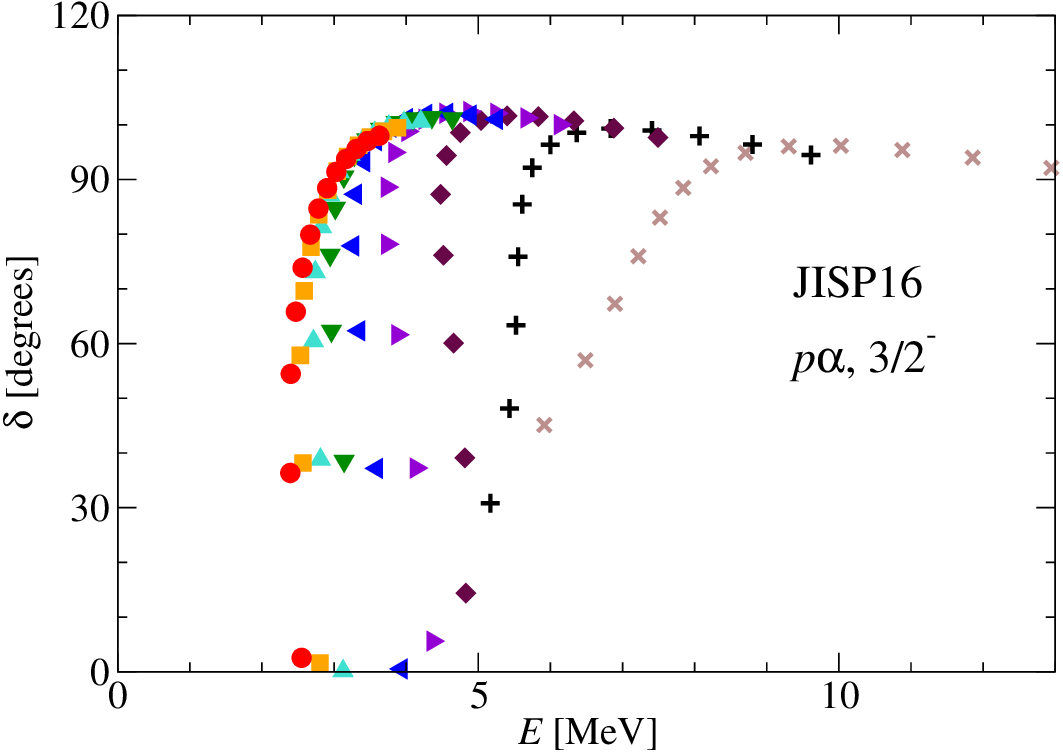}}
\caption{$p\alpha$ scattering in the $\frac32^{-}$ state with JISP16 $NN$ interaction.
The~$\frac32^{-}$ $p\alpha$ phase shifts obtained directly for all calculated $^{5}$Li 
eigenstates~$E^{(i)}_{0}$ using Eq.~\eqref{SSJM_phZ} (symbols, see Fig.~\ref{pa_p3_J_Ehw} for details).}
\label{pa_p3_J_ph_al}      
\end{figure}

Figure~\ref{pa_p3_J_Ehw} presents the results of the NCSM calculations of the
$^{5}$Li~$\frac32^{-}$ ground state  energies~$E^{(i)}_{0}$ relative to the~$p+\alpha$ threshold. The respective
phase shifts calculated using Eq.~\eqref{SSJM_phZ} for all $^{5}$Li 
eigenstates~$E^{(i)}_{0}$ are shown in the Fig.~\ref{pa_p3_J_ph_al}.

\begin{figure}[b!]
\centerline{\includegraphics[width=\columnwidth]{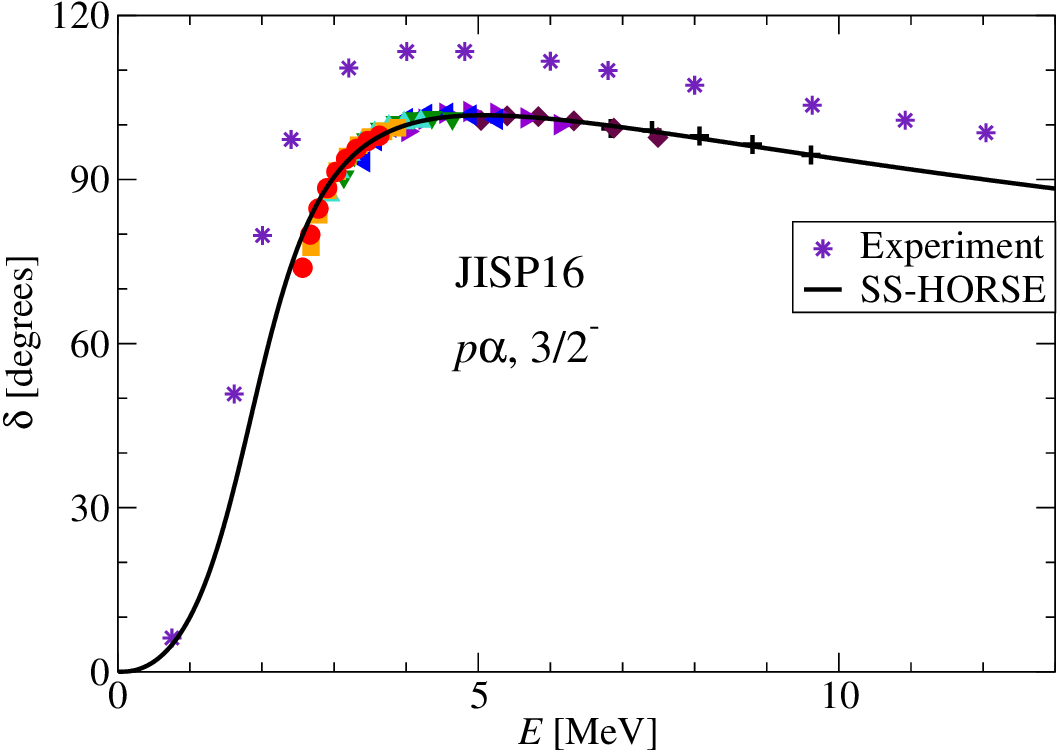}}
\caption{$p\alpha$ scattering in the $\frac32^{-}$ state with JISP16 $NN$ interaction.
The fit of the~$\frac32^{-}$ $p\alpha$ phase shifts (solid curve) and the phase shifts
obtained directly from the selected $^{5}$Li 
eigenstates~$E^{(i)}_{0}$ using Eq.~\eqref{SSJM_phZ} (symbols, see Fig.~\ref{pa_p3_J_Ehw} for details)
are compared.
 Experimental data (stars) are taken from Ref.~\cite{padata}.}
\label{pa_p3_J_phE}      
\end{figure}

\begin{figure}[t!]
\centerline{\includegraphics[width=\columnwidth]{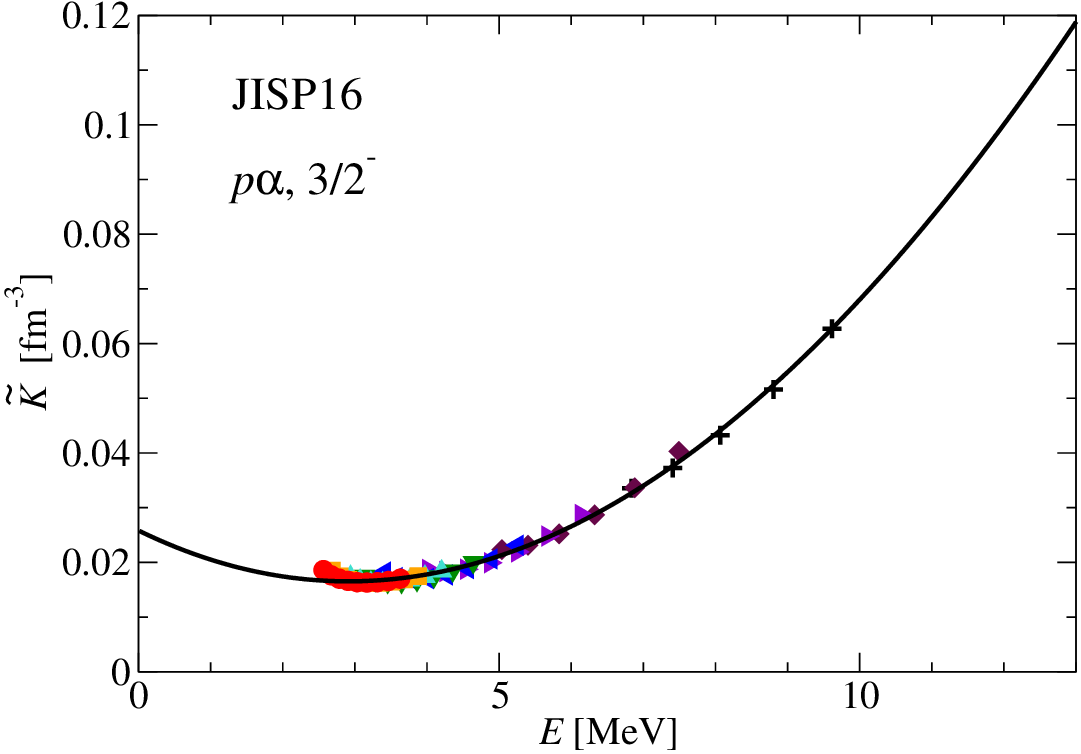}}
\caption{
Coulomb-modified  effective range function~$\widetilde{K}_{l}(E)$  
for the $p\alpha$ scattering in the $\frac32^{-}$ state with JISP16 $NN$ interaction
calculated using Eq.~\eqref{K_l} with
parameters~$w_0$, $w_1$ and~$w_2$ obtained by the fit (solid curve) and calculated using
the r.h.s.\ of Eq.~\eqref{res_K_Z_fit} at the selected eigenenergies~$E^{(i)}_{0}$ (symbols, 
see Fig.~\ref{pa_p3_J_Ehw} for details).}
\label{pa_p3_J_KE}      
\end{figure}

\begingroup
\squeezetable
\begin{table}[!b]\vspace{-3ex}
\caption{Energies~$E_r$ and widths~$\Gamma$  of resonant states~$\frac32^{-}$ and~$\frac12^{-}$  
in $^{5}$Li and $^{5}$He obtained in the NCSM-SS-HORSE approach with JISP16 and 
Daejeon16 $NN$ interactions.  The JISP16 results for $^{5}$He resonances are taken from 
Ref.~\cite{SSHORSEPRC}. $\Xi$ presents the rms deviation of energies obtained in the fit,
$D$ is the number of selected NCSM eigenenergies used in the fit. $\Delta$ is the spin-orbit splitting. 
The NCSMC results obtained with $\chi$EFT $NN$ and $NNN$ interactions  are  from 
Ref.~\cite{NCSM_RGM_PRC90} and the experimental
results are from Ref.~\cite{Hale}.}\vspace{1ex}
\label{ErG_JISP_Dj}  
\begin{ruledtabular}
\begin{tabular}{@{}c@{\hspace{1pt}}|c|c|c|c||c|c|c|c||c} 
                              & $E_r$  & $\Gamma$  &  $\Xi$   & $D$ &
                               $E_r$  & $\Gamma$   &  $\Xi$  & $D$ & $\Delta$ \\
                              &  (MeV) &  (MeV) &     (keV) & &
                                (MeV) &   (MeV) &    (keV)  &  &(MeV)
                              \\ \hline 
&\multicolumn{4}{c||}{$^5\vphantom{\int^{f}}$Li, $3/2^-$} &
\multicolumn{4}{c||}{$^5$Li, $1/2^-$}   & \\ \hline 
Experim.      & 1.69    & 1.23             &      &
     & 3.18    & 6.60             &    & &1.49  \\ \hline
JISP16 
     & 1.84  &1.80            &  43  & 60 &
           3.54  & 6.04           &  63 & 59& 1.70  \\ \hline
Daejeon16 
         & 1.52  & 1.05           &  24 & 40  &
     3.21  & 5.63          &  50 & 40 & 1.69  \\ \hline 
NCSMC   & 1.77    & 1.70 & & &
    3.11    & 7.90  & & &1.34      \\  \hline   \hline
& \multicolumn{4}{c||}{$^5\vphantom{\int^{f}}$He, $3/2^-$} &
\multicolumn{4}{c||}{$^5$He, $1/2^-$}    \\ \hline 
Experim.     & 0.80    & 0.65             &      &
      & 2.07    & 5.57             &  &  &1.27   \\ \hline
JISP16\hspace{-1pt} \cite{SSHORSEPRC}   & {0.89}  & {0.99}  & 
{70} &{68} &
 {1.86}  & {5.46}  &  {85} &{60}& {0.97}  \\ \hline 
Daejeon16 
        & 0.68  & 0.52           &  22 & 40 & 
          2.45  & 5.07           &  48&  40 & 1.77 \\ 
 \end{tabular}
\end{ruledtabular}
\end{table}
\endgroup

For the SS-HORSE analysis we should select a set of consistent (converged) NCSM
eigenstates~$E^{(i)}_{0}$ which form a single curve of the phase shifts~$\delta_l\big(E^{(i)}_{0}\big)$
vs energy as discussed in detail in Refs.~\cite{SSHORSEPRC,SSHORSE,PEPAamN,BMMS,BMMS_Z}. 
Alternatively one can use for the eigenstate selection the graph of~$E^{(i)}_{0}$ vs the scaling
parameter~$s$ or the graph of the Coulomb-modified  effective range 
function
~$\widetilde{K}_{l}\big(E^{(i)}_{0}\big)$ vs energy where the converged eigenstates should also
form a single curve. Our selection of the eigenstates~$E^{(i)}_{0}$ is illustrated by the shaded area
in 
Fig.~\ref{pa_p3_J_Ehw} while the method of the eigenstate selection is seen from
 comparing 
  Fig.~\ref{pa_p3_J_ph_al} and Fig.~\ref{pa_p3_J_phE}: the symbols in 
Fig.~\ref{pa_p3_J_ph_al} depict the phase shifts~$\delta_{1}\big(E^{(i)}_{0}\big)$ 
corresponding to all eigenstates~$E^{(i)}_{0}$ while those in 
Fig.~\ref{pa_p3_J_phE} correspond to the
selected eigenstates only. More details regarding the eigenstate selection can be found in
Refs.~\cite{SSHORSEPRC,SSHORSE} and we will follow these established procedures without further elaboration.

\strut{}A good quality reproduction of 
the Coulomb-modified  effective range function
points~$\widetilde{K}_{l}\big(E^{(i)}_{0}\big)$
by the fit is illustrated in Fig.~\ref{pa_p3_J_KE}. We note that the quality
of description  of
the functions~$\widetilde{K}_{l}(E)$ and~$K_l(E)$ by the fit in cases of other states and interactions 
is approximately the same and we shall not present the graphs of these functions in what follows.
A numerical estimate of the fit quality in our approach is the rms deviation~$\Xi$ of the 
eigenenergies~$E^{(i)}_{0}$ presented in Table~\ref{ErG_JISP_Dj}. It is seen that in all cases~$\Xi$
is of the order of few tens of~keV.

Figure~\ref{pa_p3_J_phE} demonstrates a good quality of the fit of the
phase shift points~$\delta_{1}\big(E^{(i)}_{0}\big)$. The fitted phase shifts are seen from this
panel to
reproduce qualitatively
the results of the phase shift analysis 
of the
experimental data of Ref.~\cite{padata}. However the  theoretical phase shift behavior indicates
that the resonance has a slightly higher energy and a larger width than observed experimentally;
as a result, the theoretical phase shifts lie approximately 10 degrees below those extracted from experiment at the end of the resonance region and at higher energies.

\begin{figure}[t!]
\centerline{\includegraphics[width=\columnwidth]{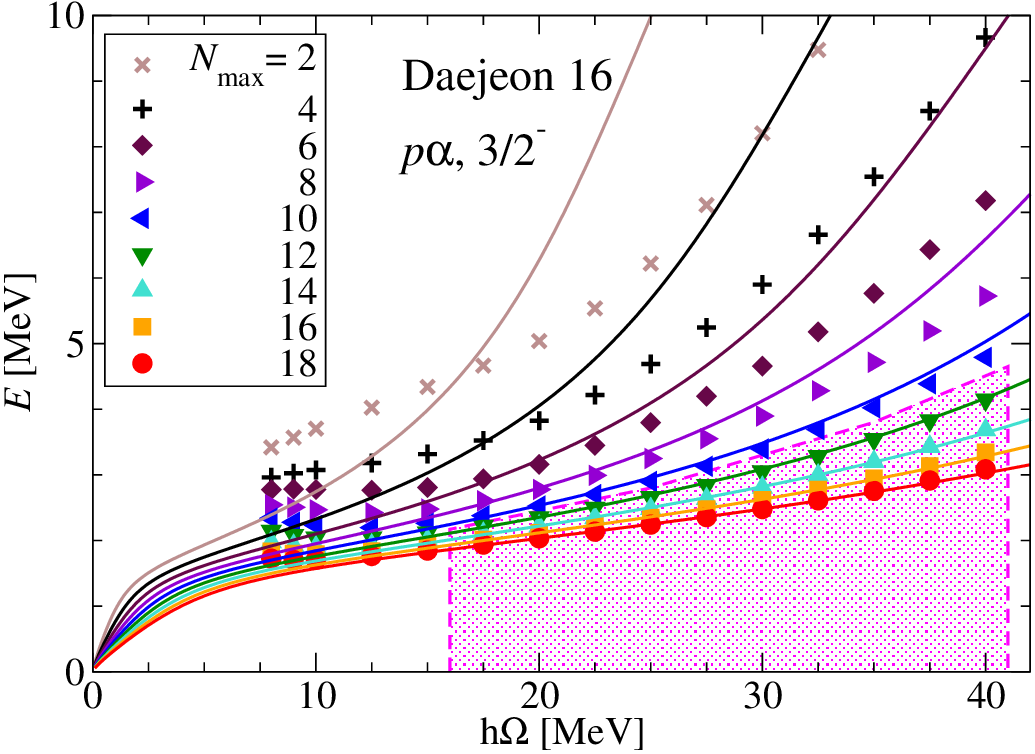}}
\caption{
The same as Fig.~\ref{pa_p3_J_Ehw} but the lowest $^{5}$Li $\frac32^{-}$ eigenenergies~$E^{(i)}_{0}$
are obtained with the Daejeon16 $NN$ interaction. }
\label{pa_p3_Dj_Ehw}      
\end{figure}
\begin{figure}[b!]
\centerline{\includegraphics[width=\columnwidth]{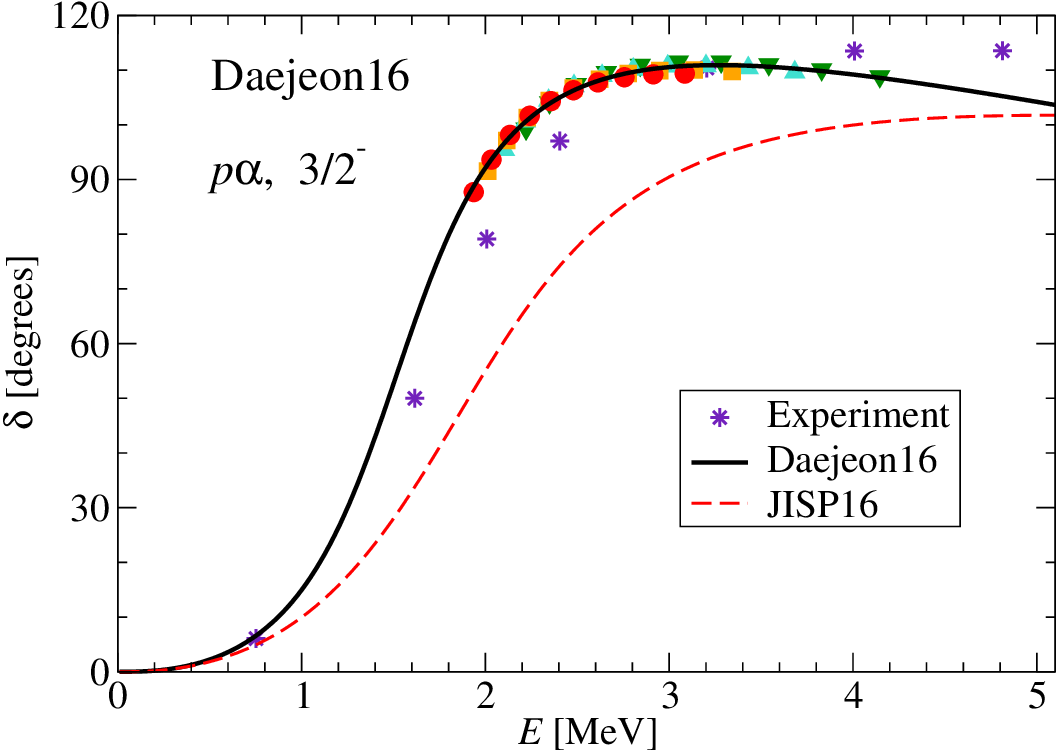}}
\caption{$p\alpha$ scattering in the $\frac32^{-}$ state with the Daejeon16 $NN$ interaction. 
Dashed curve presents the phase shifts obtained with JISP16 for comparison. 
See Fig.~\ref{pa_p3_J_phE} for other details.
}
\label{pa_p3_Dj_phE}      
\end{figure}

\begin{figure}[!]
\centerline{\includegraphics[width=\columnwidth]{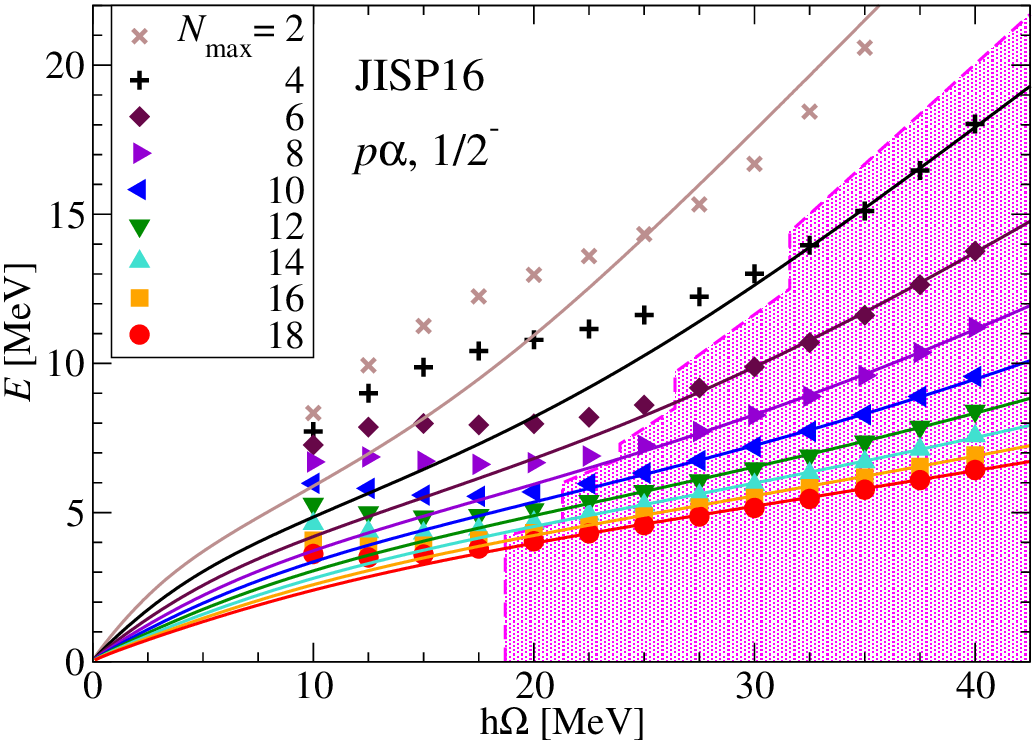}}
\caption{
The same as Fig.~\ref{pa_p3_J_Ehw} but for the lowest $^{5}$Li $\frac12^{-}$ eigenenergies~$E^{(i)}_{0}$
obtained with the JISP16 $NN$ interaction.}
\label{pa_p1_J_Ehw}
\vspace{4ex}
\centerline{\includegraphics[width=\columnwidth]{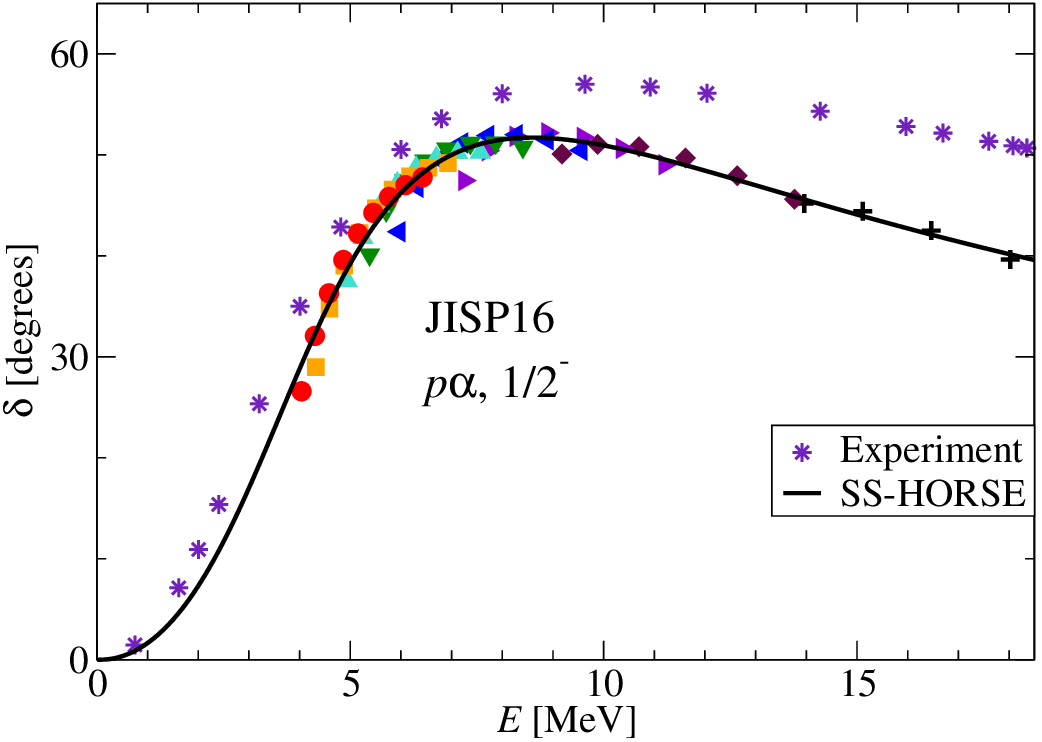}}
\caption{$p\alpha$ scattering in the $\frac12^{-}$ state with the JISP16 $NN$ interaction. See Fig.~\ref{pa_p3_J_phE} for  details.
}
\label{pa_p1_J_phE}
\vspace{4ex}
\centerline{\includegraphics[width=\columnwidth]{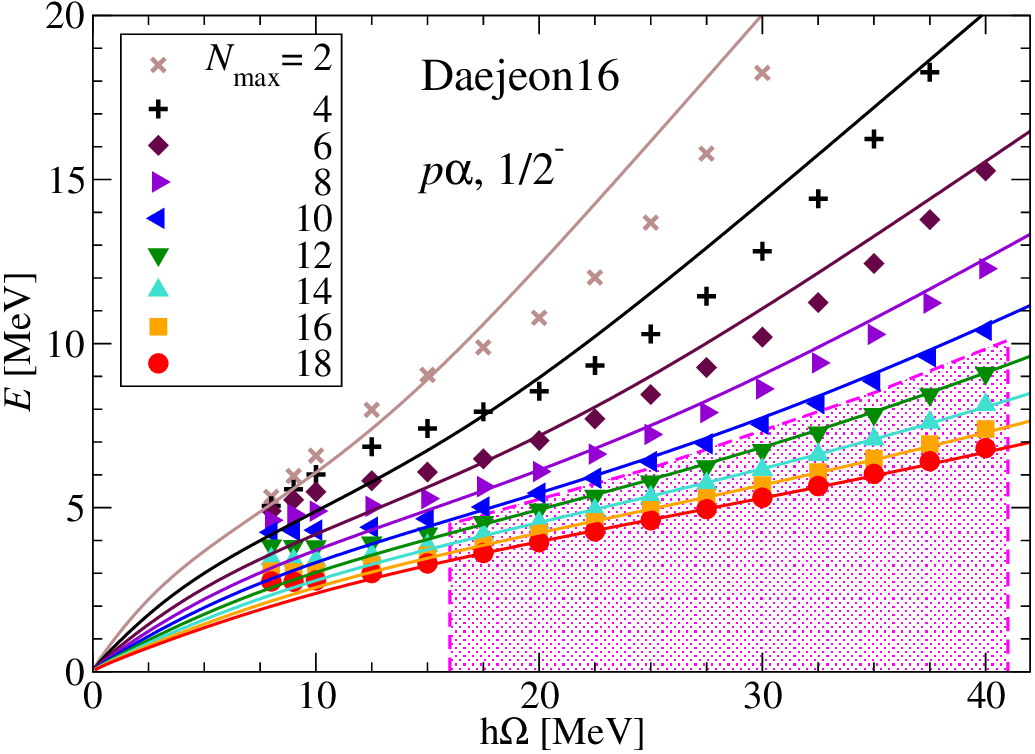}}
\caption{
The same as Fig.~\ref{pa_p3_J_Ehw} but for the lowest $^{5}$Li $\frac12^{-}$ eigenenergies~$E^{(i)}_{0}$
obtained with the Daejeon16 $NN$ interaction. 
}
\label{pa_p1_Dj_Ehw}
\end{figure}
\begin{figure}[t!]
\centerline{\includegraphics[width=\columnwidth]{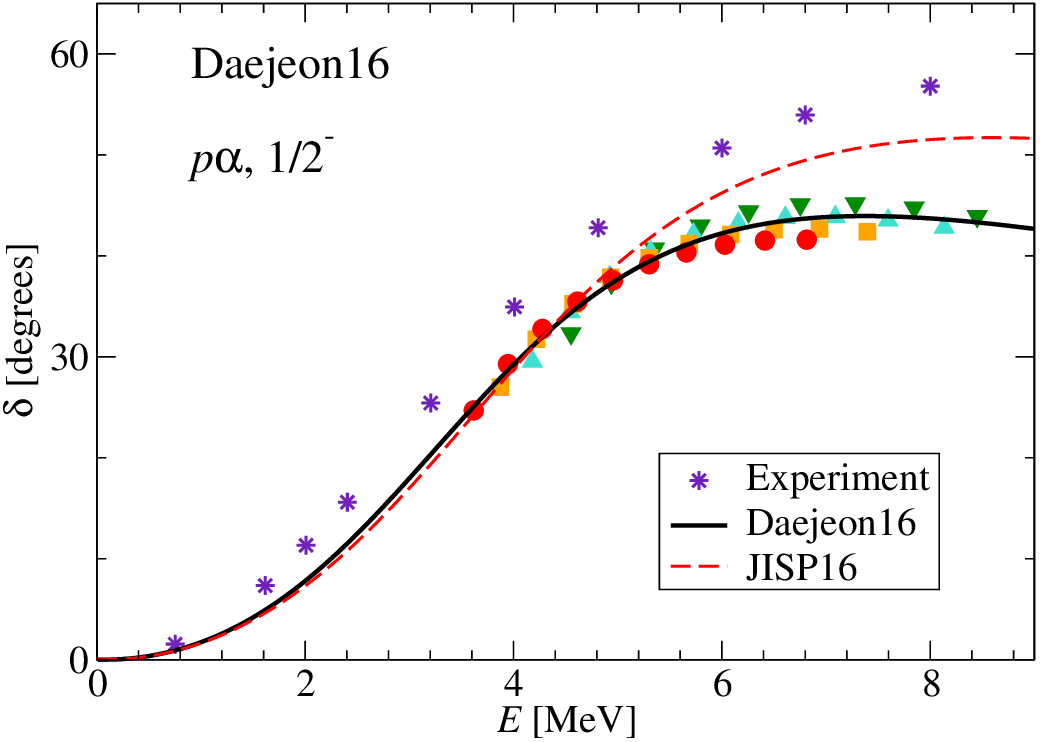}}
\caption{$p\alpha$ scattering in the $\frac12^{-}$ state with the
Daejeon16 $NN$ interaction in comparison with that obtained with JISP16. 
See 
Fig.~\ref{pa_p3_J_phE} for  details.}
\label{pa_p1_Dj_phE}
\end{figure}

The results of the calculations of the same phase shifts with the Daejeon16 $NN$ interaction are
presented in Fig.~\ref{pa_p3_Dj_Ehw} and Fig.~\ref{pa_p3_Dj_phE}. It is seen that in this case we reproduce the experimental phase shifts
in the resonance region
even better than with JISP16. However we can select for the SS-HORSE analysis  much less NCSM results than in the case of 
JISP16: only the NCSM states 
obtained with Daejeon16 with $N_{\max}\ge12$ are forming the same curve on 
the~$\delta_{1}\big(E^{(i)}_{0}\big)$ vs energy plot while in the JISP16 case we utilize 
for the SS-HORSE analysis the
results with~$N_{\max}\ge4$. In other words, surprisingly, the convergence of continuum state calculations
with the Daejeon16 $NN$ interaction is slower 
than with JISP16 while the {Daejeon16} results
in a much faster convergence of NCSM calculations for bound states of light nuclei~\cite{Daejeon16}.
The same  trends in comparing convergence of Daejeon16 and JISP16 continuum calculations 
are seen in all the rest results presented here.

\begin{figure}[b!]
\centerline{\includegraphics[width=\columnwidth]{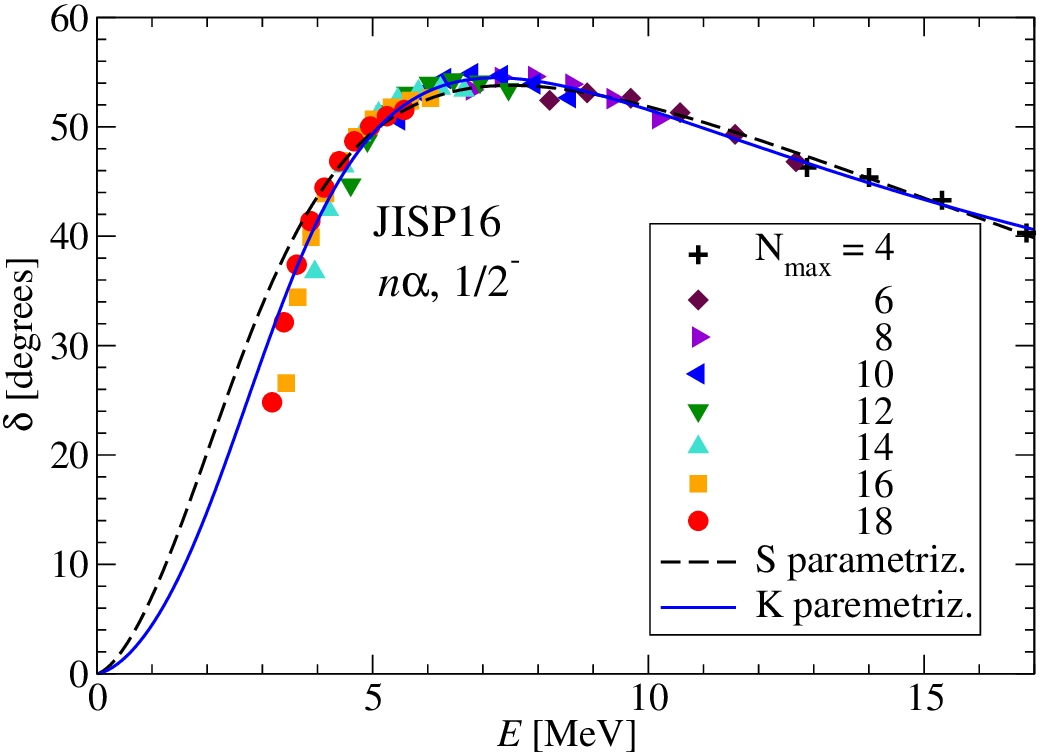}}
\caption{
The fit of the JISP16~$\frac12^{-}$ $n\alpha$ phase shifts obtained directly from the selected $^{5}$He 
eigenstates~$E^{(i)}_{0}$ using Eq.~\eqref{SSJM_phZ} (symbols, see Fig.~\ref{pa_p3_J_Ehw} for details)
using the $S$-matrix parametrization~\cite{SSHORSEPRC} (dashed curve) and 
parametrization of the effective-range function~$K_l(E)$
(solid curve).}
\label{p1_JISP_dif_S-K}
\end{figure}

The results of calculations of the $p\alpha$ scattering in the $\frac12^{-}\vphantom{_{\int_{q}}}$ state with JISP16
and Daejeon16 are presented in Figs.~\ref{pa_p1_J_Ehw}--\ref{pa_p1_Dj_phE}. 
Both interactions are reproducing well the
experimental data in the resonance region while the JISP16 phase shifts are closer to the
experiment at higher energies.


\subsection{Phase shifts of resonant \boldmath$n\alpha$ scattering}

\begin{figure}[t!]
\centerline{\includegraphics[width=\columnwidth]{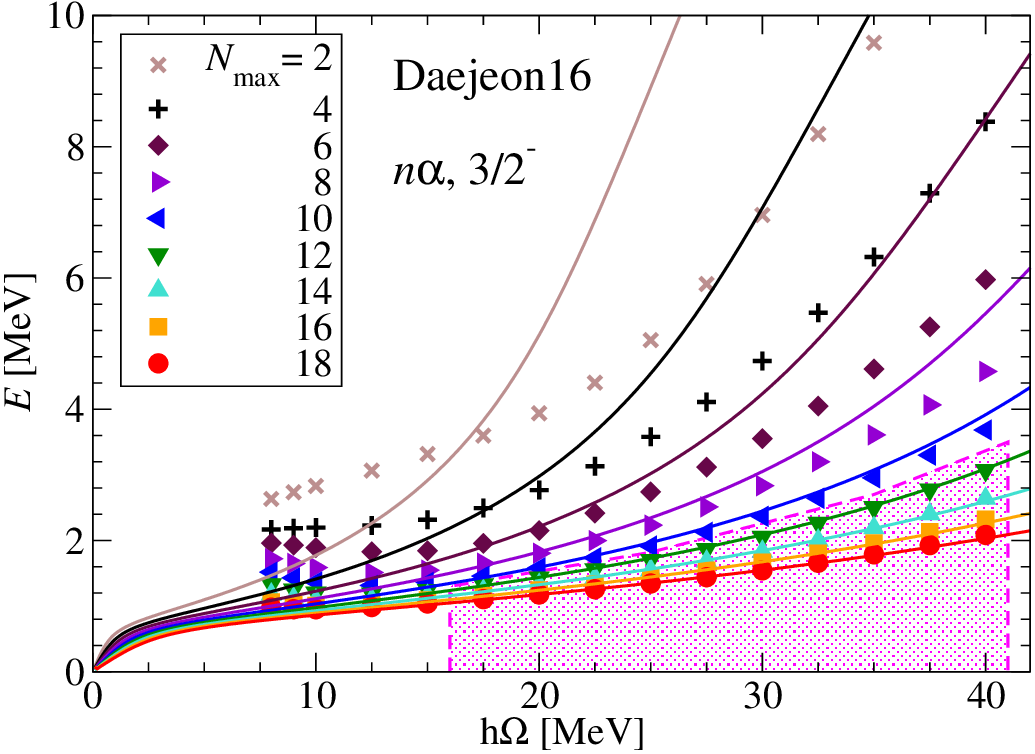}}
\caption{
The same as Fig.~\ref{pa_p3_J_Ehw} but for the lowest $^{5}$He $\frac32^{-}$ eigenenergies~$E^{(i)}_{0}$
obtained with the Daejeon16 $NN$ interaction. 
 }
\label{na_p3_Dj_Ehw}      
\end{figure}
\begin{figure}[b!]
\centerline{\includegraphics[width=\columnwidth]{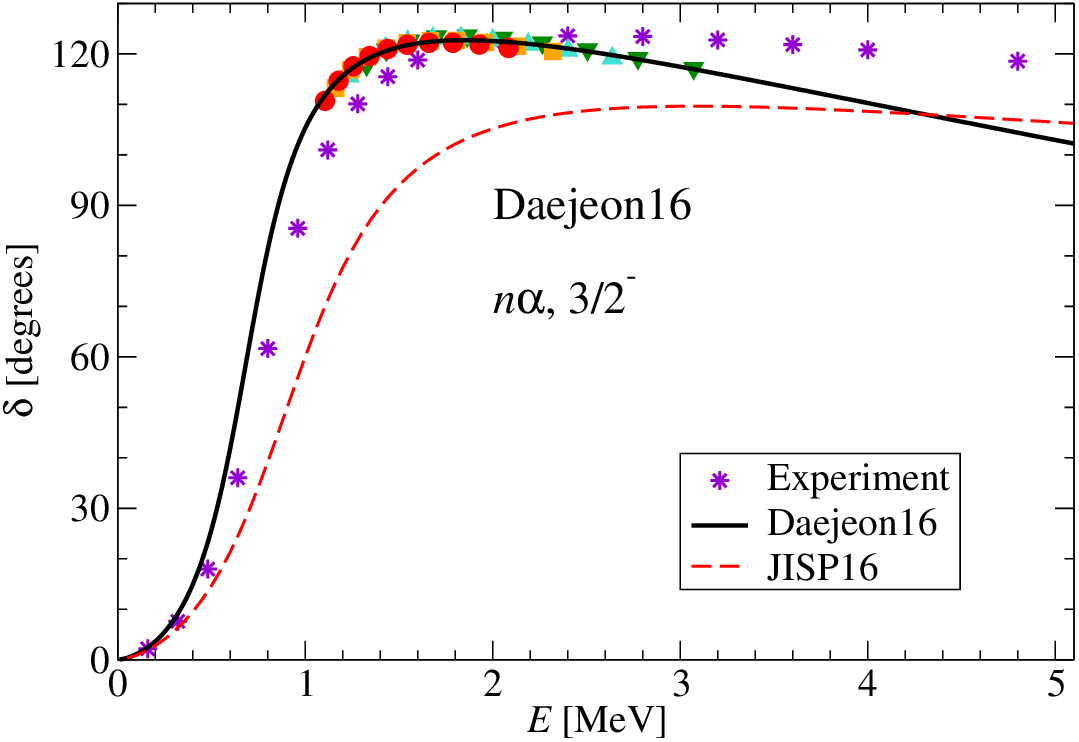}}
\caption{$n\alpha$ scattering in the $\frac32^{-}\protect\vphantom{_{\int_{q}}}$  states
with the Daejeon16  $NN$ interaction in comparison with that obtained with JISP16 in 
Ref.~\cite{SSHORSEPRC}. 
Experimental data   (stars) are taken from 
Ref.~\cite{nadata}. See 
Fig.~\ref{pa_p3_J_phE} for other details.}
\label{na_p3_Dj_phE}      
\end{figure}

We have studied the $n\alpha$ scattering within the NCSM-SS-HORSE approach with the JISP16 $NN$ 
interaction in Refs.~\cite{SSHORSEPRC,SSHORSE}. We present for completeness here the $n\alpha$ phase
shifts obtained with the Daejeon16 $NN$ interaction. We note however that the phase shifts 
and resonance parameters in 
Refs.~\cite{SSHORSEPRC,SSHORSE} were obtained using the parametrization of the
$S$-matrix in the low-energy region while here we parametrize the effective-range function~$K_l(E)$
to calculate the phase shifts and  $S$-matrix poles associated with resonances. The effective-range 
function parametrization with the same number of parameters is more accurate in describing the
phase shifts obtained directly from the NCSM eigenenergies as is seen in Fig.~\ref{p1_JISP_dif_S-K},
however this difference is pronounced in description of the wide~$\frac12^{-}$ resonance
in $^{5}$He only shown in Fig.~\ref{p1_JISP_dif_S-K}.
The difference in the phase shifts produces, of course,  the difference in the energy and width
of the~$\frac12^{-}$ resonance
in $^{5}$He  while the parameters of the narrower~$\frac32^{-}$ resonance
in $^{5}$He are only slightly affected by the different phase shift parametrizations.

\begin{figure}[t!]
\centerline{\includegraphics[width=\columnwidth]{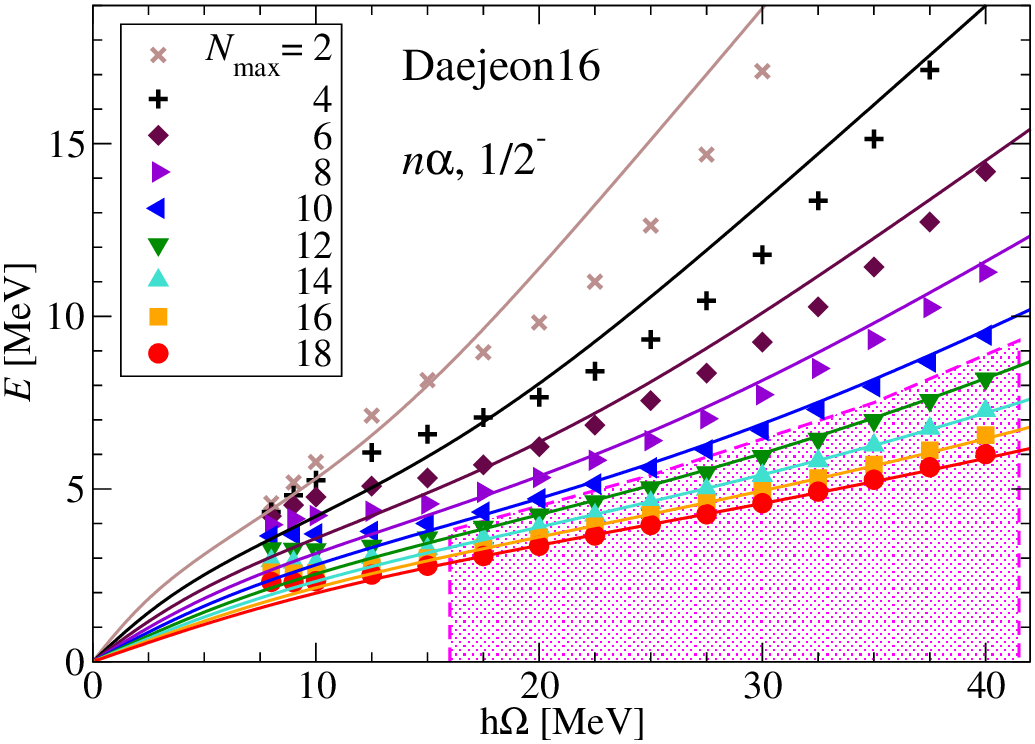}}
\caption{
The same as Fig.~\ref{pa_p3_J_Ehw} but for the lowest $^{5}$He $\frac12^{-}$ eigenenergies~$E^{(i)}_{0}$
obtained with the Daejeon16 $NN$ interaction.}
\label{na_p1_Dj_Ehw}      
\end{figure}
\begin{figure}[b!]
\centerline{\includegraphics[width=\columnwidth]{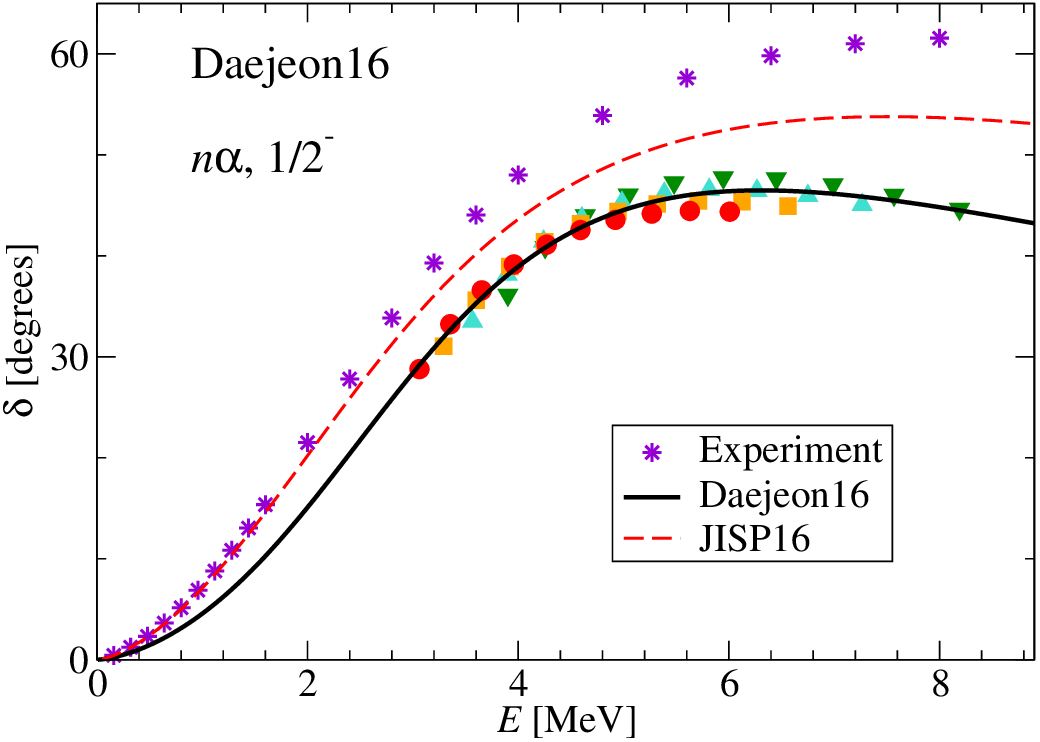}}
\caption{$n\alpha$ scattering in the $\frac12^{-}\protect\vphantom{_{\int_{q}}}$  states
with the Daejeon16  $NN$ interaction  in comparison with that obtained with JISP16 in 
Ref.~\cite{SSHORSEPRC}. 
Experimental data   (stars) are taken from 
Ref.~\cite{nadata}. See 
Fig.~\ref{pa_p3_J_phE} for other details.}
\label{na_p1_Dj_phE}      
\end{figure}

The resonant $n\alpha$ phase
shifts obtained with Daejeon16 are presented in Figs.~\ref{na_p3_Dj_Ehw}--\ref{na_p1_Dj_phE}
in comparison with those from JISP16 taken from Refs.~\cite{SSHORSEPRC,SSHORSE}.
As in the case of the $p\alpha$ scattering, the narrower $\frac32^{-}$ resonance is better described
by the Daejeon16 than by the JISP16 interaction while the~
$\frac12^{-}$ $n\alpha$ phase shifts are
reproduced better by JISP16.
We note again a faster convergence of the JISP16 calculations of $n\alpha$ scattering phase shifts
as compared with those with Daejeon16.

\subsection{\boldmath$\frac32^{-}$ and $\frac12^{-}$ resonances in $^5$Li and $^5$He nuclei}


The results for  energies and widths of the $\frac32^{-}$ and $\frac12^{-}$ resonances
in $^5$Li and $^5$He nuclei with respect to the ${\rm nucleon}+\alpha$ threshold
obtained by the numerical
location of the scattering amplitude poles as described in  Subsection~\ref{secrespole},
are presented in Table~\ref{ErG_JISP_Dj}. For comparison, we present in Table~\ref{ErG_JISP_Dj}
also the results for the $^5$Li resonances obtained with
$\chi$EFT $NN$ and $NNN$ interactions in the  {\em ab initio} NCSM/RGM approach in 
Ref.~\cite{NCSM_RGM_PRC90}. We note that the energy of the
resonance was calculated  in Ref.~\cite{NCSM_RGM_PRC90} as a position of the maximum of the 
derivative~$\frac{d\delta_{l}(E)}{dE}\vphantom{\int_{q}^{A}}$
while the resonance width was evaluated as $ \Gamma = 2/(d\delta_l/dE)|_{E=E_{r\!}}.$ The phase 
shift~$\delta_{l}(E)$ may have a contribution from a non-resonant background which can
result in some shift of the resonance energy~$E_{r}$ and in a modification of its width~$\Gamma$
in such calculations as compared with a more 
theoretically substantiated method relating the resonance parameters to the $S$-matrix
and/or scattering amplitude pole. The differences in energy and width from these different 
theoretical approaches 
may be 
large for wide resonances.

We note that all {\em ab initio} calculations of resonance parameters in $^5$Li and $^5$He nuclei
provide a good description of the experimental data of Ref.~\cite{Hale}. The difference 
in~$\frac32^{-}$ resonance energies in both nuclei obtained with different interactions is
less than~300~keV, and the experimental resonance energies are within the respective
intervals of predictions obtained
with different interactions. The theoretical predictions for the~$\frac32^{-}$ resonance widths
also embrace the experimental values. However the spread of theoretical predictions for 
the~$\frac32^{-}$ resonance width is about  750~keV in the case of $^5$Li and about~500~keV
in the case of $^5$He that appear relatively large 
compared with the widths.

In the case of the wider $\frac12^{-}$ resonances in $^5$Li and $^5$He nuclei, the spreads
of predictions for  $^5$Li also embrace the respective experimental energy and width values
while our predictions for the  $^5$He resonance energy are slightly above and, for the width,
are slightly below the experiment. However the spreads of the theoretical predictions for both
energy and width of the $\frac12^{-}$ resonances in $^5$Li and $^5$He do not exceed
approximately 600~keV 
with an exception of the NCSM/RGM
$\chi$EFT ${NN+NNN}$ prediction for the  $\frac12^{-}$ $^5$Li resonance width. Nevertheless, even the 
2.3~MeV difference
between our Daejeon16 and $\chi$EFT ${NN+NNN}$ prediction of Ref.~\cite{NCSM_RGM_PRC90} 
for the $\frac12^{-}$ $^5$Li  resonance width is much
smaller than the experimental width.  Therefore we can say that the relative accuracy of the
{\em ab initio} predictions for the $\frac12^{-}$ resonances in $^5$Li and $^5$He nuclei is
much better than that for the~$\frac32^{-}$ \mbox{resonances}. 

The difference $\Delta =\Big(\!E_r^{1/2^-}-E_r^{3/2^-}\!\Big)$ between the energies of 
the~$\frac12^{-}$ and~$\frac32^{-}$ resonances in $^5$He and $^5$Li nuclei 
{is conventionally}
associated with the spin-orbit splitting of respectively neutrons and protons in the
$p$ shell. {We note however that this interpretation should be taken with care since the 
energy difference $\Delta$ has additional contributions from the central part  of the $n{-}\alpha$ interaction potential
and from the kinetic 
energy of the relative motion of nucleon and $\alpha$ particle~\cite{Aoyama}.}
The~$\Delta$ values are presented in Table~\ref{ErG_JISP_Dj}. The
$\chi$EFT ${NN+NNN}$ interaction slightly underestimates the proton spin-orbit splitting;
the Daejeon16  overestimates both proton and neutron spin-orbit splittings while the JISP16 overestimates
the proton and underestimates the neutron spin-orbit splitting. It is interesting
to note that the differences between our predictions with JISP16 and Daejeon16 for the $^{5}$Li
resonance energies are of the order of~300~keV while  the difference in
the respective proton spin-orbit splittings~$\Delta$ is only 
about~75~keV. It is more important to note that both 
JISP16 and Daejeon16
$NN$ interactions are charge-independent; however the Daejeon16 supports nearly the 
same $p$-shell spin-orbit splittings for protons and
neutrons while the JISP16 suggests a large difference of about~800~keV between the proton and neutron
$p$-shell spin-orbit splittings which significantly 
 exceeds the 
experimental value for this difference of
approximately 200~keV.


\subsection{Non-resonant \boldmath$p\alpha$ scattering}

We have used the NCSM-SS-HORSE approach in Ref.~\cite{SSHORSEPRC,SSHORSE,PEPAamN} 
for calculations of resonant as well as non-resonant $p\alpha$ scattering. The non-resonant
phase shifts can be also calculated within the current extension of the NCSM-SS-HORSE
to the case of channels with charged colliding particles. Contrary to the phase shifts parametrizations
based on the $S$-matrix analytic properties
utilized in Refs.~\cite{SSHORSEPRC,SSHORSE,PEPAamN}, we use the same 
 Coulomb-modified effective-range function 
parametrization of Eq.~\eqref{K_l} for both resonant and non-resonant scattering.

\begin{figure}[t!]
\centerline{\includegraphics[width=\columnwidth]{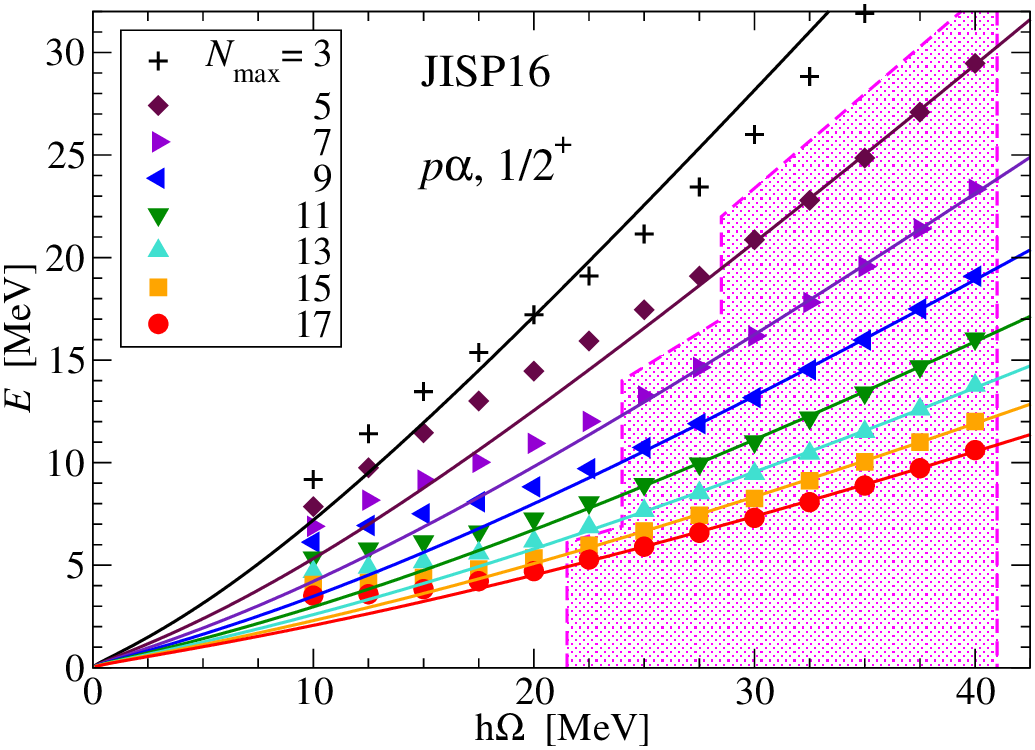}}
\caption{
The same as Fig.~\ref{pa_p3_J_Ehw} but for the lowest $^{5}$Li $\frac12^{+}$ eigenenergies~$E^{(i)}_{0}$
obtained with the JISP16 $NN$ interaction.}
\label{pa_nonres_J_Ehw}      
\end{figure}

\begin{figure}[b!]
\centerline{\includegraphics[width=\columnwidth]{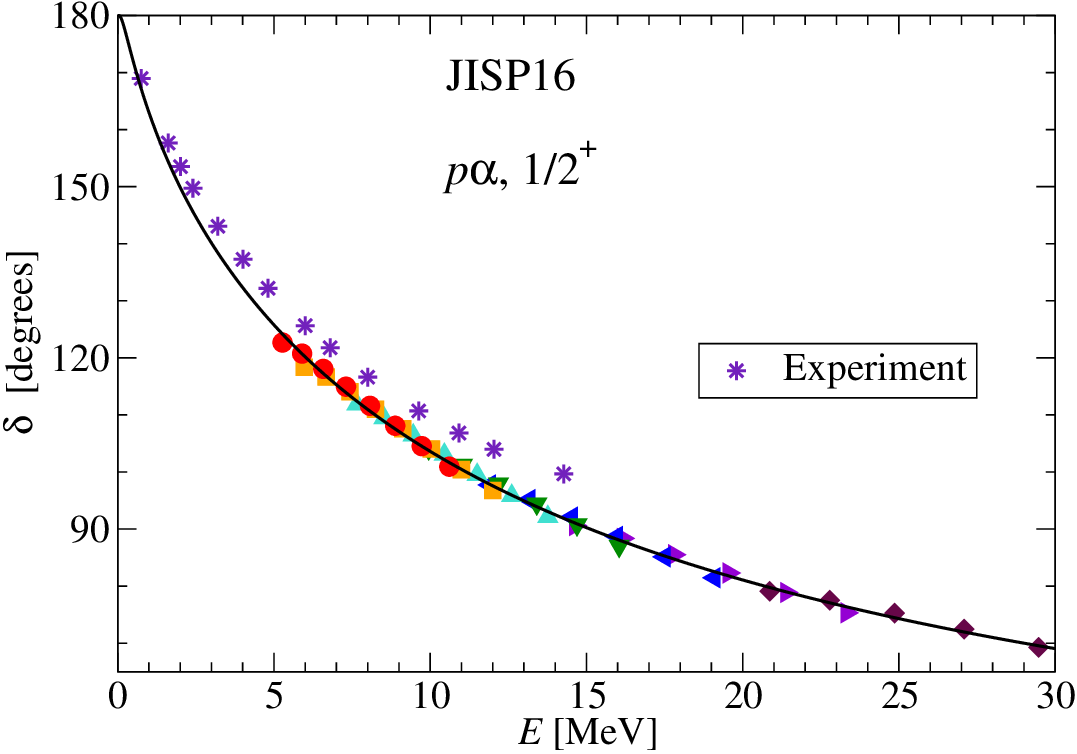}}
\caption{Non-resonant $p\alpha$ scattering in the $\frac12^{+}\protect\vphantom{_{\int_{a}}}$ state with 
Daejeon16 $NN$ interaction.
See Fig.~\ref{pa_p3_J_phE} for  details and Fig.~\ref{pa_nonres_J_Ehw} for the correspondence of 
the symbols  to the NCSM calculations 
with various~$N_{\max}$ values.}
\label{pa_nonres_J_phE}      
\end{figure}

\begin{figure}[t!]
\centerline{\includegraphics[width=\columnwidth]{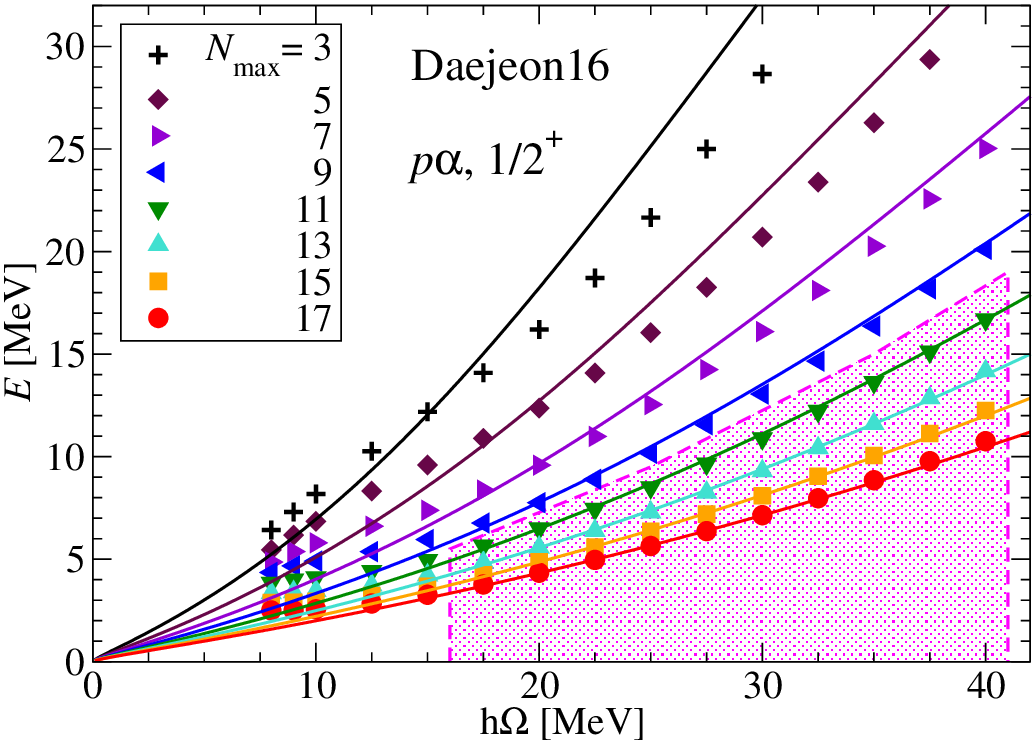}}
\caption{
The same as Fig.~\ref{pa_p3_J_Ehw} but for the lowest $^{5}$Li $\frac12^{+}$ eigenenergies~$E^{(i)}_{0}$
obtained with the Daejeon16  $NN$ interaction.}
\label{pa_nonres_Dj_Ehw}      
\end{figure}

\begin{figure}[b!]
\centerline{\includegraphics[width=\columnwidth]{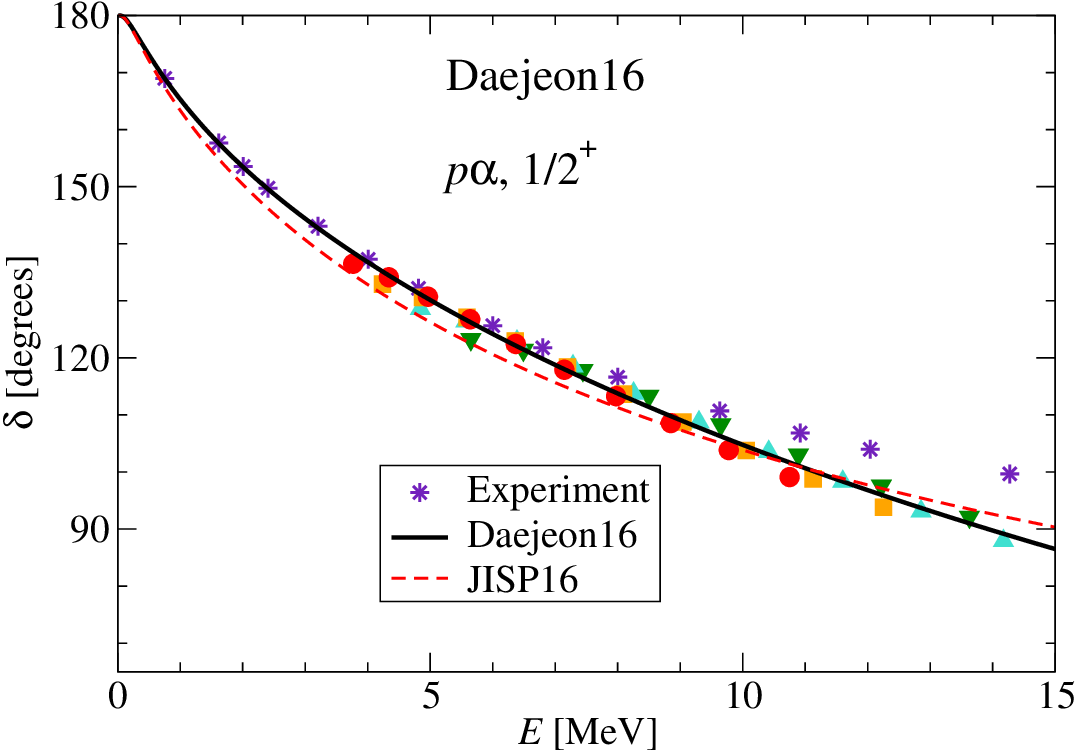}}
\caption{Non-resonant $p\alpha$ scattering in the $\frac12^{+}$ state with the
Daejeon16 $NN$ interaction in comparison with that obtained with JISP16. 
See Fig.~\ref{pa_p3_J_phE}  for  details and Fig.~\ref{pa_nonres_Dj_Ehw}  for the correspondence of 
the symbols  to the NCSM calculations 
with various~$N_{\max}$ values.}
\label{pa_nonres_Dj_phE}      
\end{figure}

The results of calculations of the non-resonant $p\alpha$ scattering phase shifts in 
the $\frac12^{+}\protect\vphantom{_{\int_{a}}}$ state with JISP16 and
Daejeon16 $NN$ interactions are presented in Figs.~\ref{pa_nonres_J_Ehw}--\ref{pa_nonres_Dj_phE}.
It is seen that JISP16
provides a faster convergence of the phase shifts in this case too. The results obtained with
JISP16 and Daejeon16 are  close to each other and reproduce well the experimental
phase shifts of Ref.~\cite{padata}.

\subsection{{Non-resonant \boldmath$n\alpha$ scattering}}

\begin{figure*}[t!]
\parbox[t]{\columnwidth}{\centerline{\includegraphics[width=\columnwidth]{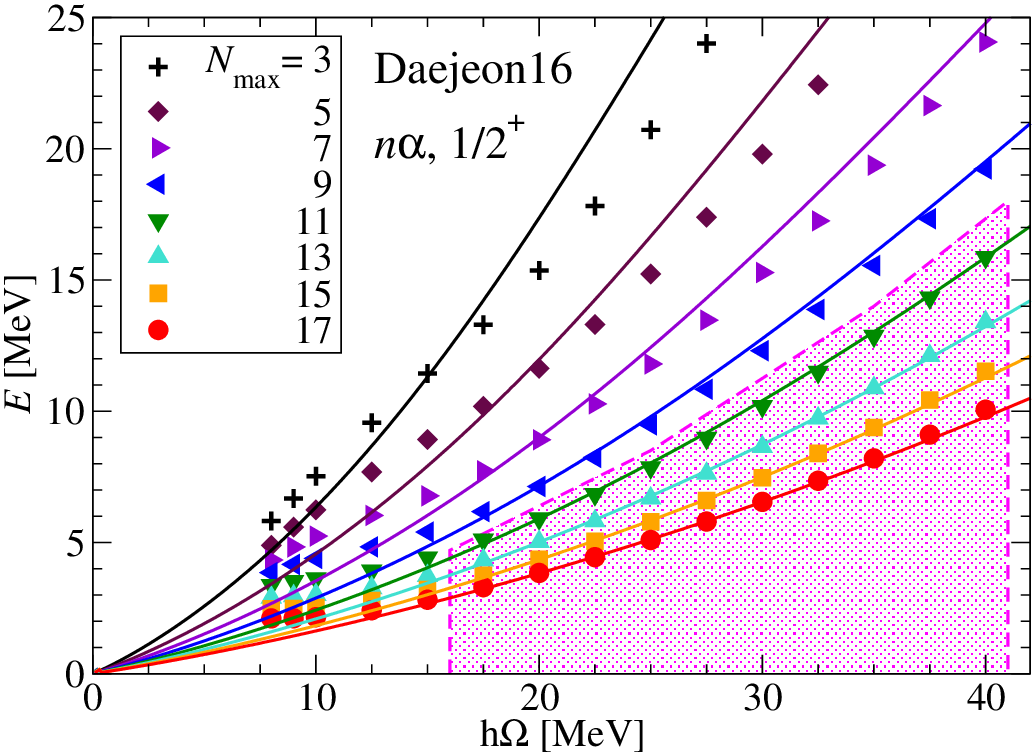}}
\caption{
The same as Fig.~\ref{pa_p3_J_Ehw} but for the lowest $^{5}$He $\frac12^{+}$ eigenenergies~$E^{(i)}_{0}$
obtained with the Daejeon16  $NN$ interaction.}
\label{na_nonres_Dj_Ehw} 
}
%
\hfill\parbox[t]{\columnwidth}{\centerline{\includegraphics[width=\columnwidth]{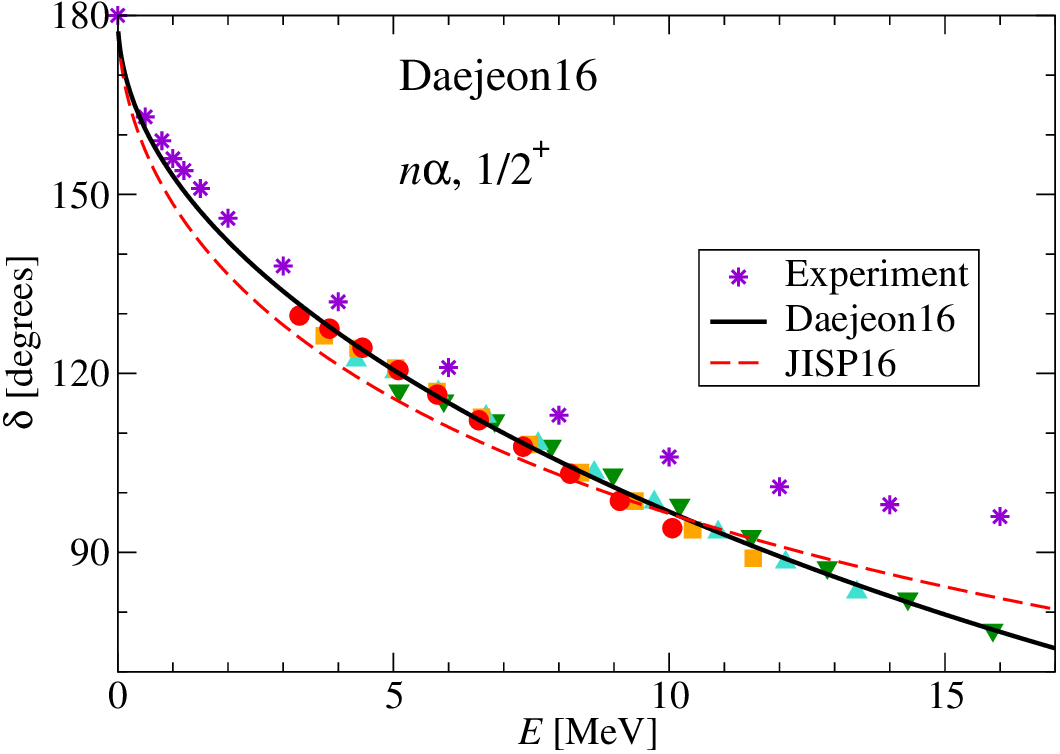}}
\caption{
Non-resonant $n\alpha$ scattering in the $\frac12^{+}$ state with the
Daejeon16 $NN$ interaction in comparison with that obtained with JISP16 in 
Ref.~\cite{SSHORSEPRC}.
Experimental data   (stars) are taken from 
Ref.~\cite{nadata}. 
See Fig.~\ref{pa_p3_J_phE} for other details and Fig.~\ref{na_nonres_Dj_Ehw}  for the correspondence of 
the symbols  to the NCSM calculations 
with various~$N_{\max}$ values.}
\label{na_nonres_Dj_phE}
}      
\end{figure*}

For completeness, we present here the results of calculations of the non-resonant $\frac12^{+}$ 
$n\alpha$ scattering phase shifts with the Daejeon16 $NN$ interaction. The results of the
NCSM calculations of the lowest~$\frac12^{+}$  $^{5}$He states
with Daejeon16 and the selection of eigenstates for the SS-HORSE
analysis is shown in Fig.~\ref{na_nonres_Dj_Ehw}; 
the obtained $\frac12^{+}$ $n\alpha$ phase shifts
are presented in Fig.~\ref{na_nonres_Dj_phE} in comparison with the respective JISP16 phase shifts
from Ref.~\cite{SSHORSEPRC} and the results of the phase-shift analysis of Ref.~\cite{nadata}.
As in the case of the non-resonant $p\alpha$ scattering, the $\frac12^{+}$ $n\alpha$ phase shifts
obtained with JISP16 and Daejeon16 $NN$ interactions are close to each other and reproduce well 
the experimental phase shifts of Ref.~\cite{nadata}.

\section{Summary}
We present here an extension of the {\em ab initio}
 NCSM-SS-HORSE approach to the case of channels
with charged colliding particles where the relative motion wave function asymptotics is 
distorted by the Coulomb interaction. The extended approach is applied to the study of $p\alpha$
scattering and resonances in the $^{5}$Li nucleus with realistic JISP16 and Daejeon16 $NN$
interactions. The analysis of the $n\alpha$ scattering and resonances in the $^{5}$He nucleus with 
the JISP16 $NN$
interaction has been performed by us  in~Refs.~\cite{SSHORSEPRC,SSHORSE,PEPAamN};
we complete this analysis here by the corresponding  calculations with Daejeon16.

We demonstrate that the extended NCSM-SS-HORSE approach works with approximately the same 
accuracy and convergence rate
as its non-extended version applicable to the channels with neutral particles. Surprisingly, we obtain
that the JISP16 interaction provides a faster convergence of the  $n\alpha$
and $p\alpha$ phase shifts than the Daejeon16
while the convergence of bound state energies in light nuclei within NCSM is much faster with 
Daejeon16 than with JISP16~\cite{Daejeon16}.

Both JISP16 and Daejeon16 provide a good \mbox{description} of the~$\frac32^{-}$ and~$\frac12^{-}$  resonances 
in $^{5}$Li and $^{5}$He nuclei as well as of the~$\frac12^{+}$ non-resonant $n\alpha$
and $p\alpha$ phase shifts. However the spin-orbit splitting of nucleons in the $p$ shell
is overestimated by the charge-independent Daejeon16 $NN$ interaction which supports nearly the
same spin-orbit splittings for neutrons and protons; the JISP16 $NN$ interaction which is also
charge-independent, overestimates the $p$-shell spin-orbit splitting for protons and underestimates
the $p$-shell spin-orbit splitting for neutrons.


\subsection*{Acknowledgments}

We are thankful to V.~D.~Efros 
and P.~Maris 
for valuable discussions.

This work  is supported in part 
by the U.S. Department of Energy 
under Grants No.~DESC00018223 (SciDAC/NUCLEI) and No.~DE-FG02-87ER40371, 
 by the Rare Isotope Science Project of 
Institute for Basic Science funded by Ministry of Science and ICT 
and 
National Research Foundation of Korea (2013M7A1A1075764). 
The development
and application of the SS-HORSE approach is supported by the
Russian  Science Foundation  under Grant No.~16-12-10048.
Computational resources were provided 
by the National Energy Research Scientific Computing Center (NERSC), which is supported by the Office 
of Science of the U.S. Department of Energy under Contract No.~DE-AC02-05CH11231, and by  
the Supercomputing Center/Korea Institute of Science and Technology Information including technical 
support (KSC-2015-C3-003).


\end{document}